\documentclass{basi}
\usepackage{astro-ph-fix}
\usepackage{txfonts}
\usepackage{graphicx}
\usepackage{dcolumn}
\newcolumntype{d}{D{.}{.}{-1}}
\def\dhead#1{\multicolumn{1}{c}{#1}}
\def\fns{\footnotesize}
\def\AD{{\sc ad}}
\def\fdg{.\!\!^\circ}
\def\itSigma{{\mathit\Sigma}}
\def\sigmaunit{W m$^{-2}$\,Hz$^{-1}$\,sr$^{-1}$}
\def\SNR(#1.#2)#3(#4.#5){{G#1${\cdot}$#2$#3$#4${\cdot}$#5}}
\def\HI{{H\,{\sc i}}}
\def\HII{{H\,{\sc ii}}}
\journal{To appear in the {\it Bulletin of the Astronomical Society of India}}
\begin{document}

\title[Galactic SNRs]{Galactic Supernova Remnants: an Updated
      Catalogue and Some Statistics}
\author[D.~A.\ Green]{D.~A.\ Green\thanks{e-mail: {\tt D.A.Green@mrao.cam.ac.uk}}\\
       Mullard Radio Astronomy Observatory, Cavendish Laboratory,
       Madingley Road, Cambridge CB3 0HE,\\ United Kingdom}

\pubyear{2004}
\volume{00}
\pagerange{\pageref{firstpage}--\pageref{lastpage}}

\setcounter{page}{1}

\date{Received 2004 September 24th} 

\maketitle

\label{firstpage}

\begin{abstract}
A catalogue of 231 Galactic supernova remnants (SNRs) is presented, and the
selection effects applicable to the identification of remnants at radio
wavelengths are discussed. In addition to missing low surface brightness
remnants, small angular size -- i.e.\ young but distant -- remnants are also
missing from the current catalogue of Galactic SNRs. Several statistical
properties of Galactic SNRs are discussed, including the
surface-brightness/diameter ($\itSigma{-}D$) relation. It is concluded that the
wide range of intrinsic properties of Galactic remnants with known distances,
together with the observational selection effects, means that use of the
$\itSigma{-}D$ relation to derive diameters and hence distances for individual
SNRs, or for statistical studies, is highly uncertain. The observed
distribution of bright SNRs, which are thought to be largely free from
selection effects, is also used to derive a simple model for the distribution
of SNRs with Galactocentric radius.
\end{abstract}


\section{Introduction}

Our Galaxy contains over two hundred known Supernova Remnants (SNRs), which are
an important source of energy and heavy element release into the interstellar
medium (ISM), and are also thought to be the sites of the acceleration of
cosmic rays. Over the last twenty years I have produced several versions of a
catalogue of Galactic SNRs, the most recent revised in 2004 January (see
Appendix~\ref{s:appendix}). Since the first version of the catalogue  was
published in Green (1984), the number of identified Galactic SNRs has increased
considerably, from 145 to 231, and here I review some of the statistical
properties of Galactic remnants based on the most recent version of the
catalogue. In Section~\ref{s:catalogue} the catalogue is described, while the
selection effects applicable to the identification of Galactic SNRs are
discussed in Section~\ref{s:selection}. Some simple statistical properties of
the remnants are presented in Section~\ref{s:simple}, with more detailed
discussions of distance-dependant statistical studies of Galactic SNRs
(including a brief discussion of some aspects of extragalactic remnants) and
the Galactic distribution of SNRs given in Sections~\ref{s:distance} and
\ref{s:distribution} respectively. The summary parameters of the 231 remnants
from the 2004 January version of the catalogue of Galactic SNRs are presented
in Appendix~\ref{s:appendix}.

\section{The Catalogue}\label{s:catalogue}

The catalogue of Galactic SNRs contains: (i) basic parameters (Galactic and
equatorial coordinates, size, type, radio flux density, spectral index, and
other names); (ii) short descriptions of the observed structure at radio, X-ray
and optical wavelengths, as applicable; (iii) other notes on distance
determinations, pulsars or point sources nearby; and (iv) references.
Appendix~\ref{s:appendix} gives the basic parameters of all 231 remnants in the
2004 January version of the catalogue, and describes these parameters in more
details. The detailed version of the catalogue is available on the
World-Wide-Web from:
 \\[6pt] \centerline{\tt http://www.mrao.cam.ac.uk/surveys/snrs/} \\[6pt]
which includes the descriptions, additional notes and references. The detailed
version is available as postscript or pdf for downloading and printing, or as
HTML web pages for each individual remnant. The web pages include links to the
`NASA Astrophysics Data System' for each of the nearly one thousand references.
Notes both on those objects no longer thought to be SNRs, and on many possible
and probable remnants that have been reported, are also included in the
detailed version of the catalogue. In addition to the observational selection
effects that are discussed further in Section~\ref{s:selection}, it should be
noted that the catalogue is far from homogeneous. Is is particularly difficult
to be uniform in terms of which objects are considered as definite remnants,
and are included in the catalogue, rather than listed as possible or probable
remnants which require further observations to clarify their nature. Although
many remnants, or possible remnants, were first identified from wide area
surveys, many others have been observed with a far from uniform set of
observational parameters, making uniform criteria for inclusion in the main
catalogue difficult. Also, some of the parameters included in the catalogue are
themselves of quite variable quality. For example, the radio flux density of
each remnant at 1~GHz. This is generally of good quality, being obtained from
several radio observations over a range of frequencies, both above and below
1~GHz. However, for a small number of remnants  -- often those which have been
identified at other than radio wavelengths -- no reliable radio flux density,
or only a limit is available (which applies to 14 remnants in the current
catalogue).

\begin{figure}
\centerline{\includegraphics[width=11.0cm]{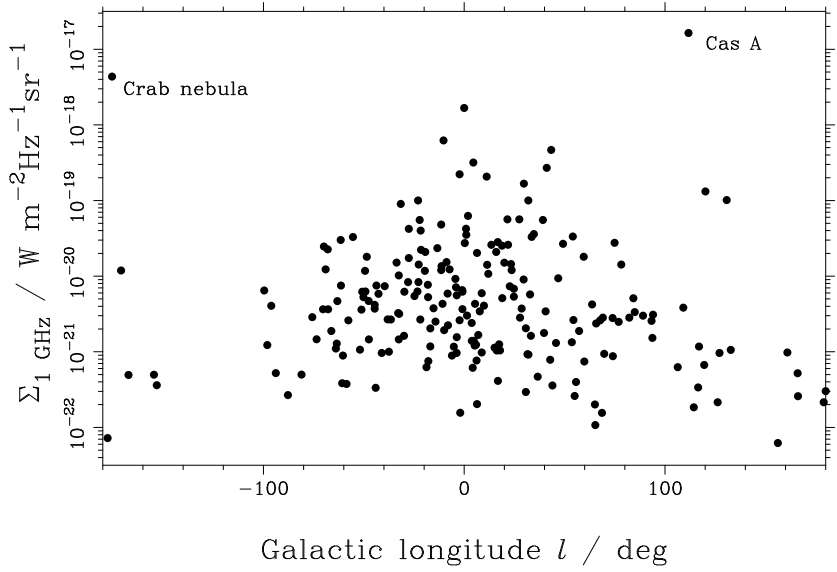}}
\caption{The distribution of surface brightness against Galactic longitude for
all 217 Galactic SNRs with defined surface brightnesses.\label{f:sigmal}}
\end{figure}

\begin{figure}
\centerline{\includegraphics[width=11.0cm]{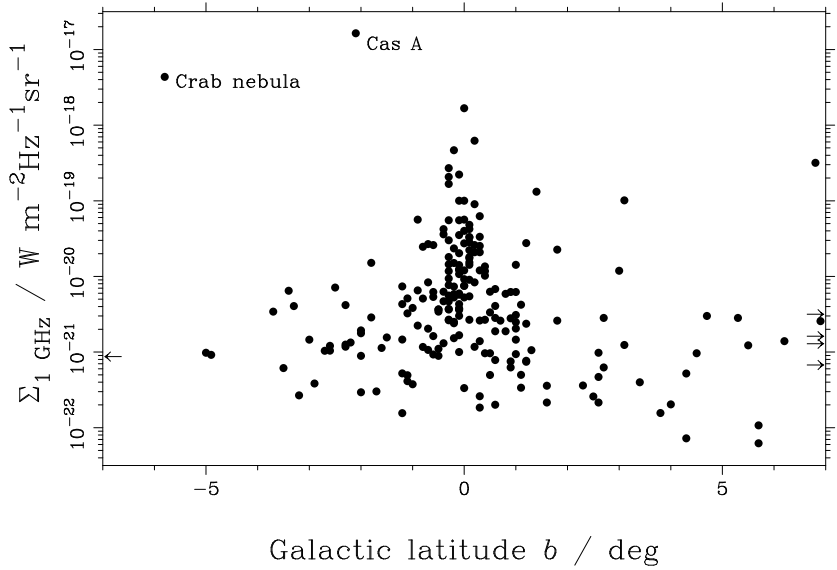}}
\caption{The distribution of surface brightness against Galactic latitude for
212 Galactic SNRs. The surface brightnesses of the five remnants with
$|b|>7^\circ$ are indicated by arrows at the left and right edges of the
plot.\label{f:sigmab}}
\end{figure}

\section{Selection Effects}\label{s:selection}

Although several Galactic SNRs have been identified at other than radio
wavelengths, in practice the dominant selection effects are those that are
applicable at radio wavelengths. Simplistically, two selection effects
apply to the identification of Galactic SNRs (e.g.\ Green 1991), due to the
difficulty in identifying (i) faint remnants and (ii) small angular size
remnants. (In the case of extragalactic SNRs, the selection effects are
different, and these are discussed briefly -- particularly in the context of
SNRs identified in M82 -- in Section~\ref{s:extragalactic}.)

\subsection{Surface Brightness}\label{s:surface}

Clearly, SNRs need to have a high enough surface brightness for them to be
distinguished from the background Galactic emission. This selection effect is
{\em not} uniform across the sky, both because the Galactic background varies
with position, and because the sensitivities of available wide area surveys
covering different portions of the Galactic plane vary. The most recent
large-scale radio surveys that have covered much of the Galactic plane are: (i)
the Effelsberg survey at 2.7~GHz (Reich et al.\ 1990; F\"urst et al.\ 1990),
which covered $358^\circ < l < 240^\circ$ and $|b| < 5^\circ$; and (ii) the
MOST survey at 843~MHz (Whiteoak \& Green 1996; Green et al.\ 1999), which
covered $245^\circ < l < 355^\circ$, but only to $|b| < 1\fdg5$.
Figs.~\ref{f:sigmal} and \ref{f:sigmab} show the distribution of surface
brightness for known SNRs against Galactic longitude and latitude. These show
that in the anti-centre and away from $b=0^\circ$, where the Galactic
background is lower, fainter remnants are relatively more common than brighter
remnants, as expected. Also, there are fewer faint remnants identified in the
4th quadrant, which is due to the narrower range of the latitude coverage of
the MOST survey compared with that of the Effelsberg survey. (There are similar
numbers of remnants with surface brightnesses less than $10^{-21}$ {\sigmaunit}
with $|b| \lesssim 1\fdg5$, in the 1st and 4th quadrant -- 9 and 7 respectively
-- but at higher Galactic latitude many more SNRs have been identified in the
1st quadrant than in the 4th -- 10 compared to 5 -- due to the wider latitude
coverage of the Effelsberg survey.)

\begin{figure}
\centerline{\includegraphics[width=11.0cm]{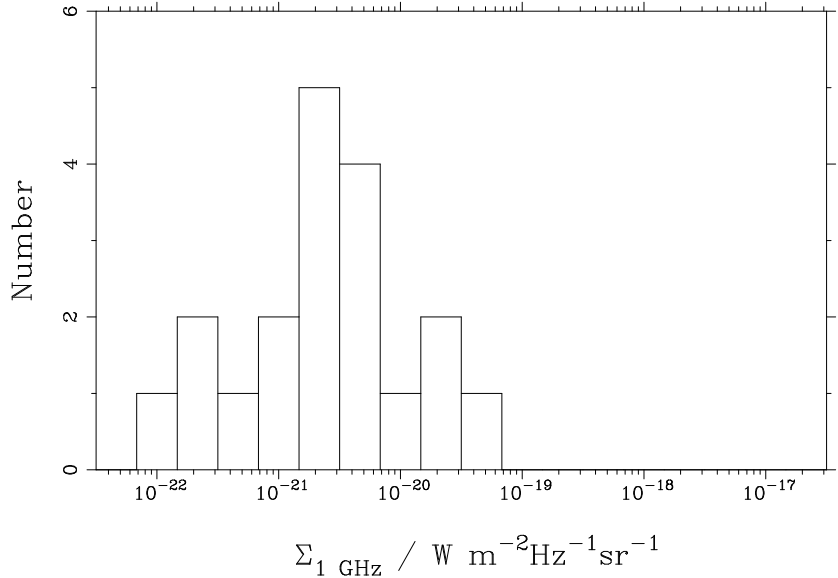}}
\caption{Histogram of the surface brightness of Galactic SNRs in the region
$358^\circ < l < 240^\circ$, $|b| < 5^\circ$ (i.e.\ the region covered by the
Effelsberg 2.7-GHz survey, Reich et al.\ 1990; F\"urst et al.\ 1990) identified
since 1991 (cf.\ Fig.~\ref{f:sigma} for the distribution for all Galactic
remnants).\label{f:post92}}
\end{figure}

The Effelsberg survey detected new SNRs down to surface brightnesses
corresponding to $\approx 2 \times 10^{-22}$ {\sigmaunit} at
1~GHz (Reich et al.\ 1988),
although the completeness limit for regions of brighter Galactic emission is
higher. This is not only because of the difficulty in identifying remnants in
the presence of extended Galactic emission, but is also due to confusion with
bright {\HII} regions (which is relatively more of a problem at higher
frequencies). Since the new SNRs identified from the Effelsberg survey were
included in the version of the SNR catalogue published in Green (1991),
consideration of the surface brightness of other remnants in the survey region
that have subsequently been identified is useful for estimating the
completeness limit for this survey. Since 1991 an additional 24 remnants within
$358^\circ < l < 240^\circ$ and $|b| < 5^\circ$ have been included in the
catalogue, most in the first quadrant. These remnants have been identified from
a variety of observations, usually covering small regions of the Galactic
plane, rather than from large area surveys. Of these, five do not have good radio
observations available, and a histogram of the surface brightnesses of the
remaining 19 is shown in Fig.~\ref{f:post92}. Of these remnants, three have
surface brightnesses above $10^{-20}$ {\sigmaunit}, two of which
(\SNR(0.3)+(0.0) and \SNR(1.0)-(0.1)) are close to the Galactic Centre, where
the background is particularly bright. Regarding these three bright remnants as
somewhat special cases, the surface brightnesses of the other more recently
identified SNRs suggest a completeness limit of $\approx 10^{-20}$
{\sigmaunit} for the Effelsberg survey.

It is difficult to estimate the completeness limit of the MOST survey in a
similar way, as only three new remnants have been identified in the MOST survey
region since the remnants identified in this survey were included in the 1996
version of the catalogue. This is due to the limited number of telescopes able
to observe this part of the Galactic plane. Of these more recently identified,
only two have surface brightnesses (of $\approx 5 \times 10^{-22}$ and $4
\times 10^{-21}$ {\sigmaunit} at 1~GHz). However, a comparison of the distribution of
the brighter SNRs in Galactic longitude suggests that the completeness limit in
the MOST survey region is not very different from that in the Effelsberg survey
region. There are 32 remnants in the 1st quadrant (i.e.\ covered by the
Effelsberg survey) with surface brightnesses above $10^{-20}$ {\sigmaunit} and
27 in the 4th quadrant (i.e.\ covered by the MOST survey). Although the
Molonglo survey covers a smaller range in Galactic latitude than the Effelsberg
survey, this difference is not important, as only one of the bright remnants in
the 1st quadrant has $|b| > 1\fdg5$,

\begin{figure}
\centerline{\includegraphics[width=11.0cm]{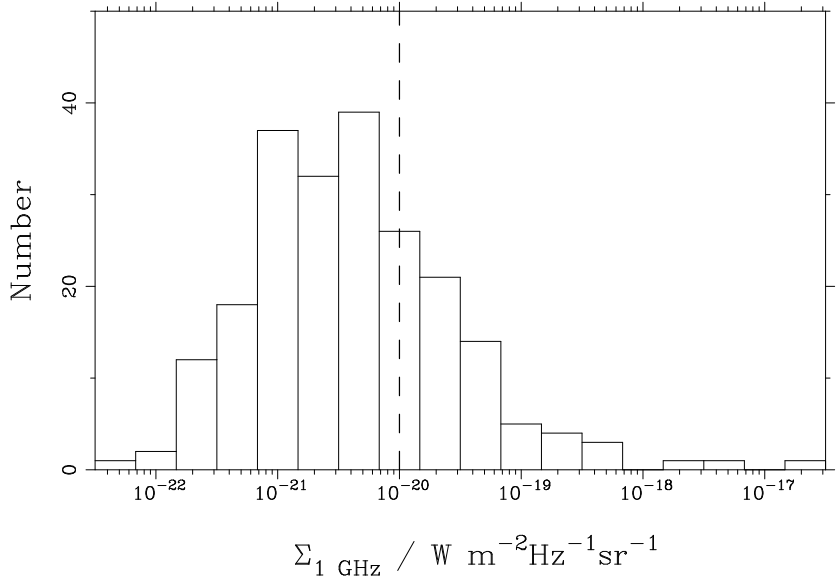}}
\caption{Distribution in surface brightness at 1~GHz of 217 Galactic SNRs. The
dashed line indicates the surface brightness completeness limit discussed in
Section~\ref{s:surface}.\label{f:sigma}}
\end{figure}

\begin{figure}
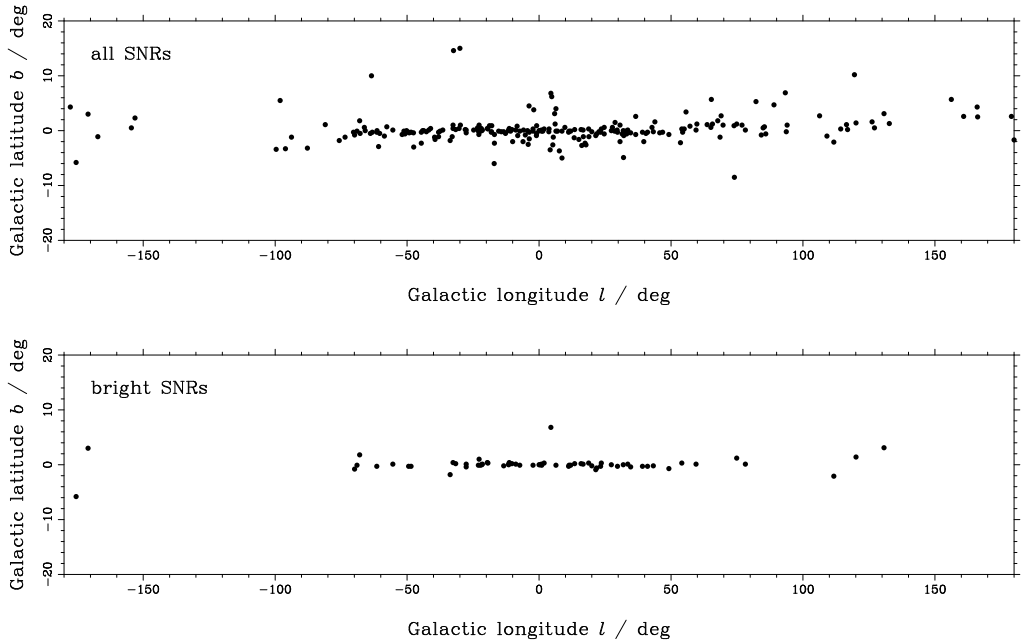

\centerline{\includegraphics[angle=270,width=13.5cm]{s-lb-all}}
\quad\\
\centerline{\includegraphics[angle=270,width=13.5cm]{s-lb-bright}}
\caption{Galactic distribution of (top) all Galactic SNR and (bottom) those
SNRs with a surface brightness at 1~GHz greater than $10^{-20}$ {\sigmaunit}.
(Note that the latitude and longitude axes are not to scale.)\label{f:lb}}
\end{figure}

So, the surface brightness limit for completeness of the current catalogue of
Galactic SNRs is approximately $10^{-20}$ {\sigmaunit}. Fig.~\ref{f:sigma}
shows a histogram of the surface brightnesses of the 217 Galactic SNRs, of
which 64 are above this nominal surface brightness limit. As noted above, many
SNRs with surface brightnesses below this limit have been identified, both from
surveys and from other observations. These fainter remnants are predominantly
in regions of the Galaxy where the background is low, i.e.\ in the 2nd and 3rd
quadrants, and away from $b=0^\circ$, as shown in Fig.~\ref{f:lb}.

It is noticeable in Fig.~\ref{f:sigmal} that there are more remnants in the 2nd
quadrant than the 3rd (21 compared with 11). It seems likely that this is due
to the fact that the 2nd quadrant is more accessible to the wider range of
northern hemisphere radio telescopes than is the 3rd quadrant. Above the
nominal surface brightness limit given above, there are only very few remnants
in the 2nd and 3rd quadrants (3 and 2 respectively), i.e.\ for these bright
remnants there is no indication of any obvious deficit of remnants in the 3rd
quadrant. At first sight, Fig.~\ref{f:sigmab} appears to show that there are
more remnants identified away from the Galactic plane at positive latitudes
than at negative latitudes. There is an asymmetry in the number of SNRs at high
Galactic latitudes. There are 11 remnants with $|b| \ge +5^\circ$, but only 4
remnants with $|b| \le -5^\circ$. Most of these high positive latitude remnants
are in the 1st and 2nd quadrants, which suggests this asymmetry is related to
Gould's Belt (e.g.\ Stothers \& Frogel 1974), which is predominantly at
positive latitudes in these quadrants. However, it is not clear that these high
latitude remnants are close enough to be associated with Gould's Belt. There is
no evidence for any asymmetry in the number of remnants at low Galactic
latitudes; there are 49 with $b \ge +1^\circ$ and 44 with $b \le -1^\circ$,
which are not statistically different.

Ongoing and future observations will no doubt continue to detect more Galactic
SNRs, although it seems very likely that most of these objects will be faint,
and hence difficult to study in detail. Currently there are several large scale
radio surveys underway that will cover much of the Galactic plane
including\footnote{Also see: {\tt http://www.ras.ucalgary.ca/IGPS/} for further
information on the first three of these surveys.}: (i) the Canadian Galactic
Plane Survey (CGPS, see: Taylor et al.\ 2003) which covers much of the northern
Galactic plane, from $l \approx 55^\circ$ to $l \approx 195^\circ$; (ii) the
VLA Galactic Plane Survey (VGPS, see: Lockman \& Stil 2004) which covers
$l=18^\circ$ to $l=67^\circ$, (iii) the Southern Galactic Plane Survey (SGPS,
see McClure-Griffiths et al.\ 2001; McClure-Griffiths 2002), and (iv) the
second-epoch Molonglo Galactic Plane Survey (MGPS2, see Green 2002), which
covers $240^\circ \le l \le 5^\circ$, $|b| \le 10^\circ$. Examples of recently
identified SNRs include two faint remnants with surface brightnesses at 1~GHz
less than a few time $10^{-22}$ {\sigmaunit}, which were found from CGPS
observations by Kothes et al.\ (2001). However, confusion with the Galactic
background -- particularly in regions near the Galactic Centre and near
$b=0^\circ$ -- will continue to be a limiting factor in the identification of
even moderately bright remnants. Comparison of radio observations over a wide
range of frequencies (so spectral index information can be used), or
observations at other than radio frequencies, may help to avoid some of the
limits caused by this confusion. Recent discoveries include three
new\footnote{These remnants are not included in the catalogue presented in
Appendix~\ref{s:appendix}, as their identification was published after that
catalogue was updated in 2004 January.}, faint SNRs near $l=11^\circ$ from
Brogan et al.\ (2004), which have surface brightness at 1~GHz of $(2{-}6)
\times 10^{-21}$ {\sigmaunit}. These new remnants were identified because of
the wide range of frequencies, and the relatively high resolution of the radio
observations.

\begin{figure}
\centerline{\includegraphics[width=11.0cm]{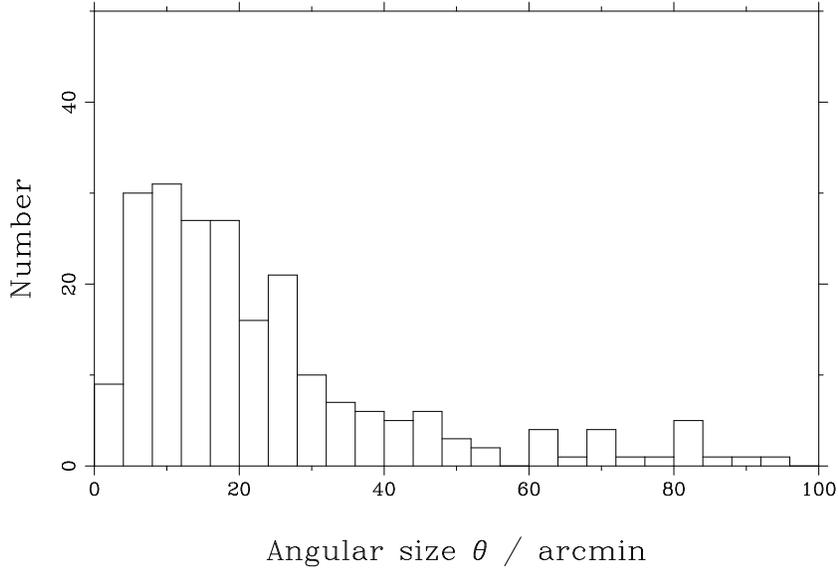}}
\caption{Histogram of the angular size of 219 Galactic SNRs (12 remnants larger
than 100 arcmin are not included).\label{f:theta}}
\end{figure}

As discussed below in Section~\ref{s:sigmad}, there is a general trend that
fainter remnants tend to be larger, and hence on average older, than brighter
remnants. However, because of the wide range of properties of Galactic SNRs
with known distances, the surface brightness selection effect applies not just
to old remnants, but also to young remnants. In particular, note that the
remnant of the SNR of {\AD} 1006 (see Table~\ref{t:historical}, below) is
fainter than the surface brightness completeness limit discussed above.

It should be noted that the study of Galactic SNRs by Filipovi\'c et al.\
(2002), which used the PMN Southern Survey images to extract flux densities of
SNRs at 5~GHz, is seriously limited by observational constraints. Since the PMN
observations were not processed to image extended objects (see Condon, Griffith
\& Wright 1993), the derived flux densities of many SNRs are in serious error.

\subsection{Angular Size}

Small angular size SNRs are likely to be missing from current catalogues. If
they are too small, then their structure is not well resolved by the available
Galactic plane surveys, and they would not be recognised as likely SNRs.
Fig.~\ref{f:theta} is shows the histogram of the angular sizes of known
remnants, which peaks at around 10~arcmin. (Note that for elongated remnants,
which have angular sizes given as  $n \times m$ arcmin$^2$ in the catalogue, a
single diameter of $\sqrt{nm}$ has been used in this histogram, and in other
figures in this paper concerned with angular size.) The limiting angular size
varies for the different available wide area surveys. As discussed above, the
radio survey that covers most of the Galactic plane is the Effelsberg 2.7-GHz
survey, which has a resolution of $\approx 4.3$~arcmin. So, for this survey, any
remnants less than about 13~arcmin in diameter (i.e.\ 3 beamwidths) are not
likely to be recognised from their structures (although, as discussed in
Section~\ref{s:missing}, some searches have been made for small remnants among
the unresolved and barely resolved sources in the Effelsberg 2.7-GHz survey).
The MOST 843-MHz survey has a much better resolution, $\approx 0.7$~arcmin,
which implies that in the region of the Galactic plane covered by this survey
only remnants smaller than about $\approx 2$ arcmin (i.e.\ 3 beamwidths) might
be expected to be missed. However, although the MOST survey
detected 18 new SNRs
(Whiteoak \& Green 1996), the smallest new remnant is \SNR(345.7)-(0.2), which is $7
\times 5$ arcmin$^2$ in extent, i.e.\ several times larger than the nominal
limit of $\approx 2$ arcmin. Thus it is difficult to quote a single angular
size selection limit for current SNR catalogues, although it is clear that it
is difficult to identify small angular size remnants from existing wide area
surveys.

This selection effect is likely to be more important for filled-centre type
remnants than for shell type remnants. Even if filled-centre remnants are large
enough to be resolved in a survey at the level of several beamwidths, their
centrally brightened structures may not be striking enough to be able to
recognise them as filled-centre remnants. Using only radio continuum
observations, it is also easy to confuse the flat-spectrum synchrotron emission
from filled-centre remnants with thermal emission. Additional observations --
e.g.\ radio polarisation (as used to identify the small angular size
filled-centre remnant \SNR(54.1)-(0.3), see Reich et al.\ 1985), radio
recombination line non-detections (see, for example, Misanovic, Cram \& Green
2002), relatively low infra-red to radio ratios (e.g.\ Cohen \& Green 2001),
X-ray emission (e.g.\ Schaudel et al.\ 2002) -- are useful to distinguish
filled-centre remnants from thermal sources.

\subsection{Missing Young but Distant SNRs}\label{s:missing}

The lack of small angular size remnants -- i.e.\ young but distant remnants --
is particularly clear when the remnants of known `historical' Galactic
supernovae (see Stephenson \& Green 2002) are considered. These remnants are
relatively close-by -- as is expected, since their parent SNe were seen
historically -- and therefore sample only a small fraction of the Galactic
disc. Consequently we expect many more similar, but more distant remnants in
our Galaxy (e.g.\ Green 1985), but these are not present in current catalogues.

\begin{table}
\caption{Parameters of known historical SNRs, plus Cas~A.\label{t:historical}}
\begin{center}
\small\tabcolsep 2pt
\smallskip
\begin{tabular}{cccdcdccdccd}\hline
           &                    &          & \multicolumn{3}{c}{as observed}                     & & \multicolumn{2}{c}{if at 8.5~kpc} & & \multicolumn{2}{c}{if at 17~kpc} \\ \cline{4-6}\cline{8-9}\cline{11-12}
    date   & name or            & distance &   \dhead{size}   & $\itSigma_{\rm 1~GHz}$ & \dhead{$S_{\rm 1~GHz}$}& &   size    & \dhead{$S_{\rm 1~GHz}$} & &   size    & \dhead{$S_{\rm 1~GHz}$} \\
           & remnant            &   /kpc   &  \dhead{/arcmin} &   {\fns /{\sigmaunit}} &       \dhead{/Jy}      & &  /arcmin  &      \dhead{/Jy}        & &  /arcmin  &       \dhead{/Jy}       \\ \hline
     --    & Cas A              &   3.4    &     5     & $1.6 \times 10^{-17}$ &     2720        & &    2.0     &      435             & &    1.0    &   109                \\
{\AD} 1604 & Kepler's           &   2.9    &     3     & $3.2 \times 10^{-19}$ &       19        & &    1.0     &        2.2           & &    0.5    &     0.55             \\
{\AD} 1572 & Tycho's            &   2.3    &     8     & $1.3 \times 10^{-19}$ &       56        & &    2.3     &        4.1           & &    1.1    &     1.0              \\
{\AD} 1181 & 3C58               &   3.2    &     7     & $1.0 \times 10^{-19}$ &       33        & &    2.6     &        4.7           & &    1.3    &     1.2              \\
{\AD} 1054 & {\fns Crab nebula} &   1.9    &     6     & $4.4 \times 10^{-18}$ &     1040        & &    1.4     &       52             & &    0.7    &    13                \\
{\AD} 1006 & {\fns \SNR(327.6)+(14.6)} &   2.2    &    30     & $3.2 \times 10^{-21}$ &       19        & &    7.7     &        1.3           & &    3.9    &     0.31             \\ \hline
\end{tabular}
\end{center}
\end{table}

Table~\ref{t:historical} gives the distances, angular sizes, flux
densities and surface brightnesses at 1~GHz, for the remnants of known
historical supernovae from the last thousand years, plus Cas~A (which although
its progenitor was not seen -- so it is not strictly a historical remnant -- is
known to be only about 300 years old). The distances used for these remnants
are those given in Section~\ref{s:distances}. This table also lists the
parameters of these remnants when scaled to larger distances of 8.5 and 17~kpc,
i.e.\ to represent how they would appear if they if they were at the other
side of the Galaxy (from the Galactic Centre, to the far point on the Solar
Circle). The number of other young (i.e.\ less than a thousand year old) SNRs
expected in the Galaxy can be estimated in two simple ways:
%
%
%
(i) from the expected supernova rate of one every 45 to 70 years (Cappellaro
2003), 15 to 22 young remnants are expected in total;
(ii) considering the fraction of the Galactic disc sampled by the historical
supernovae (6 in a thousand years), which are within $\approx 4$~kpc, i.e.\
about 16 per cent of the Galactic disc modelled as being uniform and having a
radius of $\approx 10$~kpc, implies there should be $\approx 40$ young remnants
(see also the discussions in Strom 1994). Of these, $\approx 80\%$ (see
Cappellaro 2003) are expected to be the remnants of massive supernovae -- i.e.\
from type Ib/Ic/II SNe -- and therefore be close to the Galactic plane.

\begin{figure}
\centerline{\includegraphics[width=6cm]{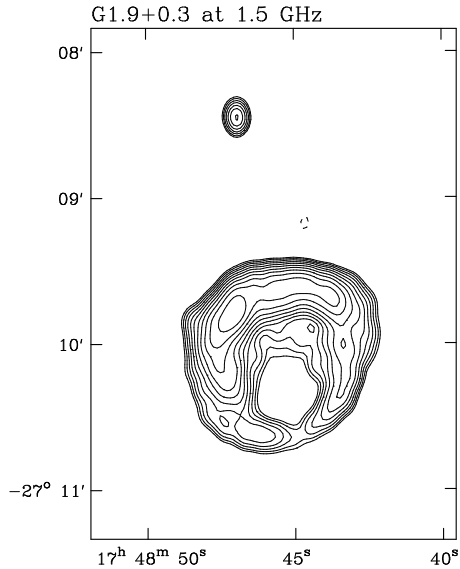}}
\caption{A VLA image of \SNR(1.9)+(0.3) at 1.5~GHz, with a resolution of $9.4
\times 7.2$ arcsec$^2$ at a position angle of $3^\circ$. The contour levels are
at $-1, 1 \sqrt{2^n}$ mJy beam$^{-1}$ for $n=0, 1, 2\dots$ (with the negative
contour dashed). The coordinates are J2000.0.\label{f:g1.9}}
\end{figure}

\begin{figure}
\centerline{\includegraphics[width=11.0cm]{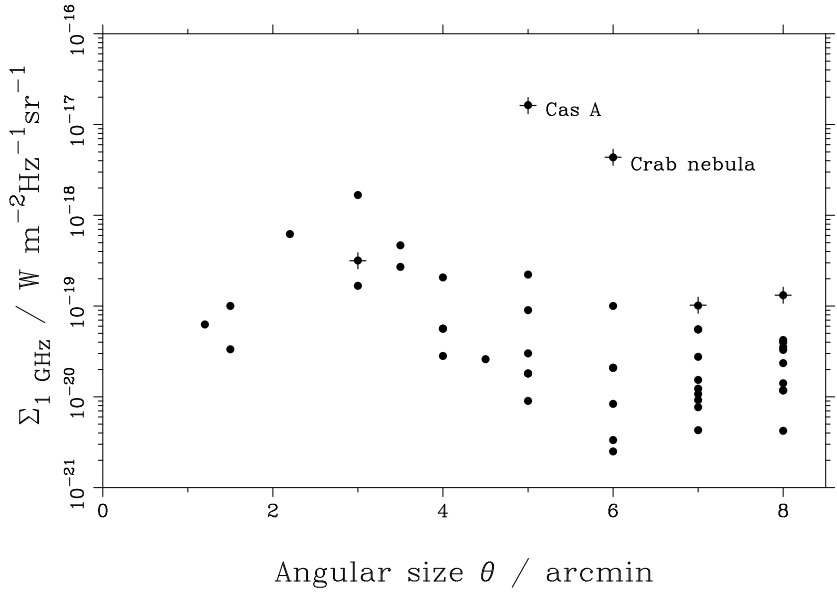}}
\caption{Distribution in surface brightness at 1~GHz against angular size for
known Galactic SNRs of angular size $\le 8$ arcmin. The five historical
remnants from Table~\ref{t:historical} included in this plot are indicated by
additional crosses. (For all but the smallest few remnants in the catalogue,
the angular size is given to the nearest arcmin.)\label{f:sigmatheta}}
\end{figure}

From Table~\ref{t:historical}, any young SNRs in the Galaxy similar to the
known historical remnants, but in the far half of the Galaxy, would generally
be expected to have angular sizes less than a few arcmin, high surface
brightness, greater than $\approx 10^{-19}$ {\sigmaunit} (although remnants
similar to the remnant of the SN of {\AD} 1006 would be much fainter). These
remnants would also be expected to lie close to the Galactic plane, with $|b|
\lesssim 1^\circ$. Although the above estimates are rather uncertain due to
intrinsic Poisson uncertainties from the small numbers of known historical
remnants, they imply that about a dozen or more young but distant remnants
might be expected. However,  there are very few such remnants in the current Galactic
SNR catalogue. In fact there are only 3 known remnants with angular sizes of 2
arcmin or less: \SNR(1.9)+(0.3), \SNR(54.1)+(0.3) and \SNR(337.0)-(0.1).
Of these, \SNR(1.9)+(0.3) is the smallest, with an angular size of only
1.2~arcmin. No distance measurement is available for this remnant -- which
being close to $l=0^\circ$ makes kinematic methods unreliable -- but even if it
were at the far side of the Galaxy, at say 17~kpc, its physical size would only
be 6~pc. This is comparable to the sizes of the known historical remnants in
Table~\ref{t:historical} (which have physical diameters of 5, 3, 5, 7, 3 and 19
pc respectively). Another indication that this is indeed a young remnant is
that it shows a circularly symmetric limb-brightened shell of radio emission
(see Fig.~\ref{f:g1.9} for a previously unpublished image of this remnant at
1.5~GHz, from 1985 observations made with the NRAO's VLA\footnote{The National
Radio Astronomy Observatory is a facility of the National Science Foundation
operated under cooperative agreement by Associated Universities, Inc.}). The
known young shell remnants all show highly circular structures, whereas older
remnants tend to be less circular, as is expected as they expand into differing
regions of the interstellar medium. The distance to the filled-centre remnant
\SNR(54.1)+(0.3) is uncertain (e.g.\ Camilo et al.\ 2002), although its small
angular size of 1.5~arcmin again suggests it is physically small, and therefore
young, even if it were to be situated at the edge of the Galaxy. A reasonably
accurate distance estimate is available for \SNR(337.0)-(0.1) (see
Section~\ref{s:distances}), and this is $\approx 11$~kpc, which is consistent
with this being a young but distant SNR (with a diameter of $\approx 5$~pc,
from its angular size of 1.5 arcmin). The deficit of small remnants is also
illustrated in Fig.~\ref{f:sigmatheta}, which shows there are very few known
Galactic remnants with high surface brightnesses and small angular sizes.

This deficit of young but distant remnants has long been recognised, but
searches for remnants of this type (e.g.\
Green
\& Gull 1984; Helfand et al.\ 1984; Green 1985, 1989; Sramek et al.\ 1992; Misanovic et al.\
2002; see also Saikia et al.\ 2004) have had only limited success (identifying
the small remnants \SNR(1.9)+(0.3) and \SNR(54.1)+(0.3) noted above). Since the
missing young but distant remnants are expected to have angular sizes of a few
arcmin or less, they will not have been resolved sufficiently by single-dish
radio surveys with resolutions of several arcminutes, so will not have been
recognised as SNRs. Searching for these remnants is not easy because there are
very many candidate sources in single-dish surveys to choose from (e.g.\ in the
1st quadrant, there are over a thousand compact sources in the Effelsberg
survey with $|b| \le 1^\circ$), and only a fraction of these have been observed
with high enough resolution to be able to identify them if they were small
angular size SNRs. Moreover, the missing young but distant remnants are likely
to be in the most complex, and hence most confused, regions of the Galactic plane,
as being distant they will be close to $b=0^\circ$. The use of additional
observational indicators (e.g.\ radio spectral index, infra-red to radio ratios) has yet not proved efficient for improving such searches, nor have the
MOST survey (see Whiteoak \& Green 1996) and the NVSS (see Condon et al.\
1998), which cover the Galactic plane with slightly higher resolution than
available single-dish surveys. (Note, however, that the NVSS does contain
previously unrecognised SNRs, for example \SNR(353.9)-(2.0), with an angular
size of 13 arcmin, see Green 2001a.) The fact that such missing, small remnants
are likely to be in complex regions of the plane may mean that confusion is a
very significant problem, and not just at radio wavelengths. Further searches
for these missing young but distant remnants are required.

It should be noted that there are unlikely to be other luminous remnants in the
Galaxy like Cas A and the Crab nebula. Any such remnants, even on the far side
of the Galaxy, would have relatively high flux densities, and the nature of
all such sources in the Galactic plane is known. On the other hand, the remnant
of the SN of {\AD} 1006 is faint -- possibly because it is far from the
Galactic plane, in a low density region -- and distant remnants similar to this
would be particularly difficult to detect, as they would have both a small
angular size and low surface brightness. However, any remnants similar to the
other three historical remnants are detectable.

\section{Some Simple SNR Statistics}\label{s:simple}

In the current version of the catalogue, 77\% of remnants are classed as shell
(or possible shell), 12\% are composite (or possible composite), and 4\% are
filled-centre (or possible filled centre) remnants. The remaining 7\% have not
yet been observed well enough to be sure of their type, or else are objects
which are conventionally regarded as SNRs although they do not fit well into
any of the conventional types (e.g.\ CTB80 ($=$\SNR(69.0)+(2.7)), MSH
17$-$3{\em 9} ($=$\SNR(357.7)-(0.1))). Since the 1991 version of the catalogue
(Green 1991), the proportion of shell remnants in the catalogue has stayed very
similar, with the proportion of composite remnants increasing from 8\%, and the
proportion of filled centre remnants has decreasing from 7\%. The increase in
the proportion of composite remnants is because more recent, improved
observations have continued to identify many more shell remnants, but have also
identified faint, pulsar powered nebulae in what until then had been identified
as pure shell remnants (e.g.\ W44 ($=$\SNR(24.7)-(0.4))), and also that faint
shells have been detected around some filled-centre remnants (e.g.\
\SNR(21.5)-(0.9)).

\begin{figure}
\centerline{\includegraphics[width=11.0cm]{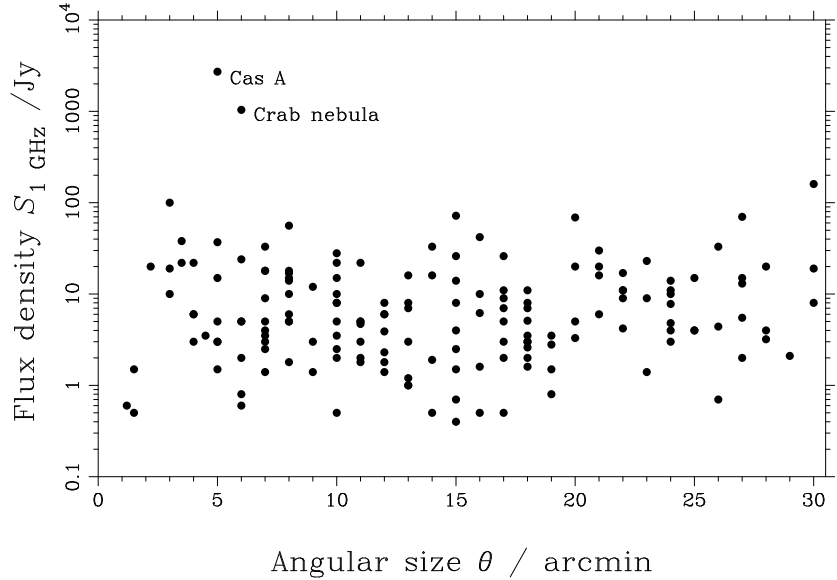}}
\caption{Distribution of flux density at 1~GHz against angular size for known
Galactic SNRs with diameters $\le 30$~arcmin.\label{f:stheta}}
\end{figure}

There are 14 Galactic SNRs that are either not detected at radio wavelengths,
or are poorly defined by current radio observations, so that their flux
density at 1~GHz cannot be determined with any confidence: i.e.\ 94\% have a
flux density at 1~GHz included in the catalogue. Of the catalogued remnants,
36\% are detected in X-ray, and 23\% in the optical. At both these wavelengths,
Galactic absorption hampers the detection of distant remnants.

Some of the properties of the Galactic SNRs in the catalogue, which are not
shown in other sections of this paper, are shown in Fig.~\ref{f:stheta}. This
shows the flux density at 1~GHz versus angular size for SNRs less than 30
arcmin in extent. This is of interest in terms of which remnants may appear as
bright, relatively compact Galactic plane sources (e.g.\ in future Planck
surveys). The most prominent sources are, not surprisingly, the very bright
SNRs Cas~A and the Crab nebula (see Table~\ref{t:historical}), which have
similar angular sizes, and very high flux densities. Cas~A has the higher flux
density at 1~GHz by a factor of about 2.6, but because the Crab nebula has a
much flatter spectrum (with $\alpha \approx 0.30$ compared with $\approx 0.77$
for Cas~A, e.g.\ Baars et al.\ 1977), the Crab nebula has the higher flux
density at frequencies above about 8~GHz. The statistics of the radio spectral
indices of Galactic SNRs are not discussed here, although there is a short
discussion of these in Green (2001b).

\section{Distance Dependant SNR Statistics}\label{s:distance}

\subsection{Distances to SNRs}\label{s:distances}

Many studies of Galactic SNRs require knowledge of the distances to remnants
(or equivalently their physical sizes, since their angular sizes are known).
However, accurate distances are not available for many known SNRs. The
distances that are available are obtained from a wide variety of methods --
e.g.\ optical expansion and proper motion studies, 21-cm {\HI} absorption
spectra, {\HI} column density (see Foster \& Routledge 2003), association with
{\HI} or CO features in the surrounding interstellar medium, or association
with other objects -- each of which is subject to their own uncertainties, and
some of which are subjective. Table~\ref{t:distances} presents distances for 47
Galactic SNRs available in the literature.
(In a few cases distances estimates to SNRs are also available from the
distances to associated pulsar, derived from the observed pulsar dispersion
measure and a model of the Galactic electron density distribution. However,
these have not been used in Table~\ref{t:distances}.)
In several cases the distances derived from {\HI} absorption measurements have
been recalculated using a modern `flat' rotation curve with a Galactocentric
radius of 8.5~kpc and a constant rotation speed of 220 km s$^{-1}$.
Additionally, in a few cases the distances given depend on re-interpretation of
the published observations. For \SNR(11.2)-(0.3) and \SNR(21.5)-(0.9), the
distances given correspond to the near distances of the last strong {\HI}
absorption seen (see also Safi-Harb et al.\ 2001 for \SNR(21.5)-(0.9)).


\begin{table}
\caption{Galactic SNRs with distance measurements or
estimates.\label{t:distances}}
\begin{center}
\smallskip
\footnotesize
\begin{tabular}{rdllr}\hline
   remnant         & \dhead{distance} & method                       & reference                    & notes \\
                   &   \dhead{/kpc}   &                              &                              &       \\ \hline
   \SNR(4.5)+(6.8)  &  2.9 & optical proper motion/velocity & Blair et al.\ (1991)             &     \\
   \SNR(6.4)-(0.1)  &  1.9 & {\HI} absorption               & Vel{\'a}zquez et al.\ (2002)     &     \\
  \SNR(11.2)-(0.3)  &  4.4 & {\HI} absorption               & Becker et al.\ (1985)            & a,b \\
  \SNR(18.8)+(0.3)  & 14.0 & association with CO            & Dubner et al.\ (2004)            &     \\
  \SNR(21.5)-(0.9)  &  4.6 & {\HI} absorption               & Davelaar et al.\ (1986)          & a,b \\[3pt]
  \SNR(27.4)+(0.0)  &  6.8 & {\HI} absorption               & Sanbonmatsu \& Helfand (1992)       &     \\
  \SNR(33.6)+(0.1)  &  7.8 & {\HI} absorption               & Frail \& Clifton (1989)          & a   \\
  \SNR(34.7)-(0.4)  &  2.8 & {\HI} absorption               & Caswell et al.\  (1975)          & a   \\
  \SNR(39.7)-(2.0)  &  3.0 & association with {\HI}         & Dubner et al.\ (1998)            &     \\
  \SNR(43.3)-(0.2)  & 10.0 & association with {\HI}         & Lockhart \& Goss (1978)          & a   \\[3pt]
  \SNR(49.2)-(0.7)  &  6.0 & association with CO            & Koo et al.\ (1995)               &     \\
  \SNR(53.6)-(2.2)  &  2.8 & association with {\HI}         & Giacani et al.\ (1998)           &     \\
  \SNR(55.0)+(0.3)  & 14.0 & association with {\HI}         & Matthews et al.\ (1998)          &     \\
  \SNR(74.0)-(8.5)  &  0.4 & optical proper motion/velocity & Blair et al.\ (1999)             &     \\
  \SNR(74.9)+(1.2)  &  6.1 & {\HI} column density           & Kothes et al.\ (2003)            &     \\[3pt]
  \SNR(84.2)-(0.8)  &  4.5 & association with CO            & Feldt \& Green (1993)            &     \\
  \SNR(89.0)+(4.7)  &  0.8 & association with CO and {\HI}  & Tatematsu et al.\ (1990)         &     \\
  \SNR(93.3)+(6.9)  &  2.2 & {\HI} column density           & Foster \& Routledge (2003)       &     \\
  \SNR(93.7)-(0.2)  &  1.5 & association with {\HI}         & Uyan{\i}ker et al.\ (2002)       &     \\
 \SNR(109.1)-(1.0)  &  3.0 & association with {\HII} region & Kothes et al.\ (2002)            &     \\[3pt]
 \SNR(111.7)-(2.1)  &  3.4 & optical proper motion/velocity & Reed et al.\ (1995)              &     \\
 \SNR(114.3)+(0.3)  &  0.7 & association with {\HI}         & Yar-Uyan{\i}ker et al (2004)     &     \\
 \SNR(116.5)+(1.1)  &  1.6 & association with {\HI}         & Yar-Uyan{\i}ker et al (2004)     &     \\
 \SNR(116.9)+(0.2)  &  1.6 & association with {\HI}         & Yar-Uyan{\i}ker et al (2004)     &     \\
 \SNR(119.5)+(10.2) &  1.4 & association with {\HI}         & Pineault et al.\ (1993)          &     \\[3pt]
 \SNR(120.1)+(1.4)  &  2.3 & optical proper motion/velocity & Chevalier et al.\ (1980)         &     \\
 \SNR(130.7)+(3.1)  &  3.2 & {\HI} absorption               & Roberts et al.\ (1993)           &     \\
 \SNR(132.7)+(1.3)  &  2.2 & association with CO            & Routledge et al.\ (1991)         &     \\
 \SNR(166.0)+(4.3)  &  4.5 & association with {\HI}         & Landecker et al.\ (1989)         &     \\
 \SNR(166.2)+(2.5)  &  8.0 & association with {\HI}         & Routledge et al.\ (1986)         &     \\[3pt]
 \SNR(184.6)-(5.8)  &  1.9 & various                        & Trimble (1973)                   &     \\
 \SNR(189.1)+(3.0)  &  1.5 & optical absorption             & Welsh \& Sallmen (2003)          &     \\
 \SNR(205.5)+(0.5)  &  1.6 & various                        & Odegard (1986)                   &     \\
 \SNR(260.4)-(3.4)  &  2.2 & association with {\HI}         & Reynoso et al.\ (1995)           &     \\
 \SNR(263.9)-(3.3)  &  0.3 & pulsar parallax                & Caraveo et al.\ (2001)           &     \\[3pt]
 \SNR(292.0)+(1.8)  &  6.0 & various                        & Gaensler \& Wallace (2003)       &     \\
 \SNR(292.2)-(0.5)  &  8.4 & {\HI} absorption               & Caswell et al.\ (2004)           &     \\
 \SNR(296.8)-(0.3)  &  9.6 & association with {\HI}         & Gaensler et al.\ (1998a)         &     \\
 \SNR(315.4)-(2.3)  &  2.3 & optical velocity               & Sollerman et al.\ (2003)         &     \\
 \SNR(320.4)-(1.2)  &  5.2 & {\HI} absorption               & Gaensler et al.\ (1999)          &     \\[3pt]
 \SNR(327.4)+(0.4)  &  4.8 & {\HI} absorption               & McClure-Griffiths et al.\ (2001) &     \\
 \SNR(327.6)+(14.6) &  2.2 & optical proper motion/velocity & Winkler et al.\ (2003)           &     \\
 \SNR(332.4)-(0.4)  &  3.1 & {\HI} absorption               & Caswell et al.\ (1975)           & a   \\
 \SNR(337.0)-(0.1)  & 11.0 & various                        & Sarma et al.\ (1997)             &     \\
 \SNR(348.5)+(0.1)  &  8.0 & {\HI} absorption               & Caswell et al.\ (1975)           & a   \\[3pt]
 \SNR(348.7)+(0.4)  &  8.0 & {\HI} absorption               & Caswell et al.\ (1975)           & a   \\
 \SNR(349.7)+(0.2)  & 14.8 & {\HI} absorption               & Caswell et al.\ (1975)           & a   \\ \hline
\end{tabular}\\
\noindent Notes: a) distance recalculated using modern rotation curve; b) see
text for further discussion.
\end{center}
\end{table}

\begin{figure}
\centerline{\includegraphics[width=11.0cm]{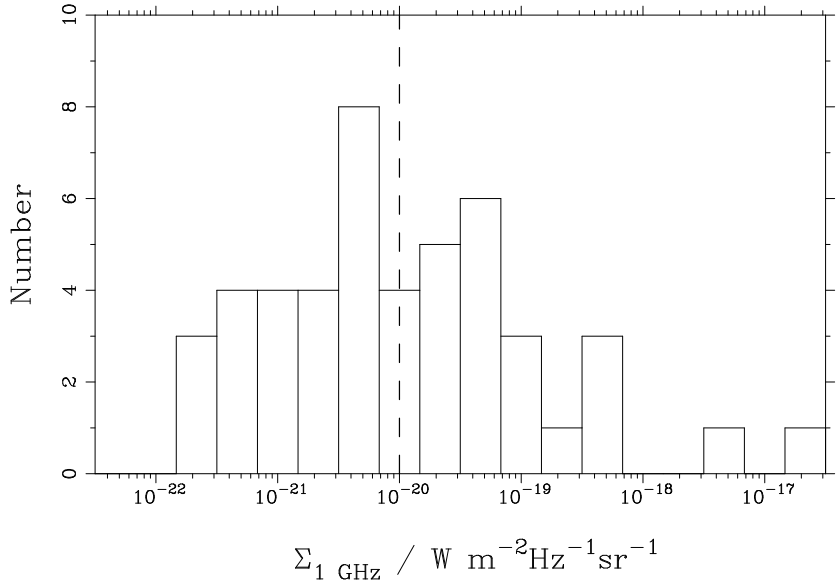}}
\caption{Distribution in surface brightness at 1~GHz of 47 Galactic SNRs with
known distances (see Section~\ref{s:distances}). The dashed line indicates the
surface brightness completeness limit discussed in
Section~\ref{s:surface}.\label{f:withd}}
\end{figure}

The uncertainties in these distances are far from uniform. For kinematic
distances -- which are a large majority of the distances given in
Table~\ref{t:distances} -- there are always some uncertainties in deriving
distances from observed velocities, due to deviations from circular motion
(especially an issue for nearby remnants, and for those near $l=0^\circ$ and
$180^\circ$ where the observed velocity does not depend strongly on distance)
and ambiguities inside the Solar Circle. Generally, the published errors in
kinematic distances are less than $\approx 25\%$, although there are additional
possible larger uncertainties depending on whether the (often subjective)
association of a particular feature with a remnant is actually correct. A
potential bias for statistical studies is that many distance methods are more
easily applied to brighter remnants than to fainter ones. This is particularly the
case for 21-cm {\HI} absorption studies, which depend on a radio continuum from
the remnant being bright, otherwise any absorption could not be studied in
reasonable detail. But this also applies to some of the other methods, e.g.\
association with other {\HI} or CO features in the ISM, where fainter remnants
will not be well defined, so that clear morphological association with other
features will be more difficult. Indeed, Fig.~\ref{f:withd} shows a histogram
of the surface brightness of Galactic SNRs with known distances, which shows
these tend to be the brighter Galactic SNRs overall (see Fig.~\ref{f:sigma}). Thus,
it is likely that, at a given diameter, SNRs with known distances are biased to
brighter remnants. The number of SNRs with available distances is sufficiently
large that statistical studies -- see Section~\ref{s:sigmad} -- show that the
range of intrinsic luminosities of Galactic SNRs is large.

\begin{figure}
\centerline{\includegraphics[angle=270,width=9cm]{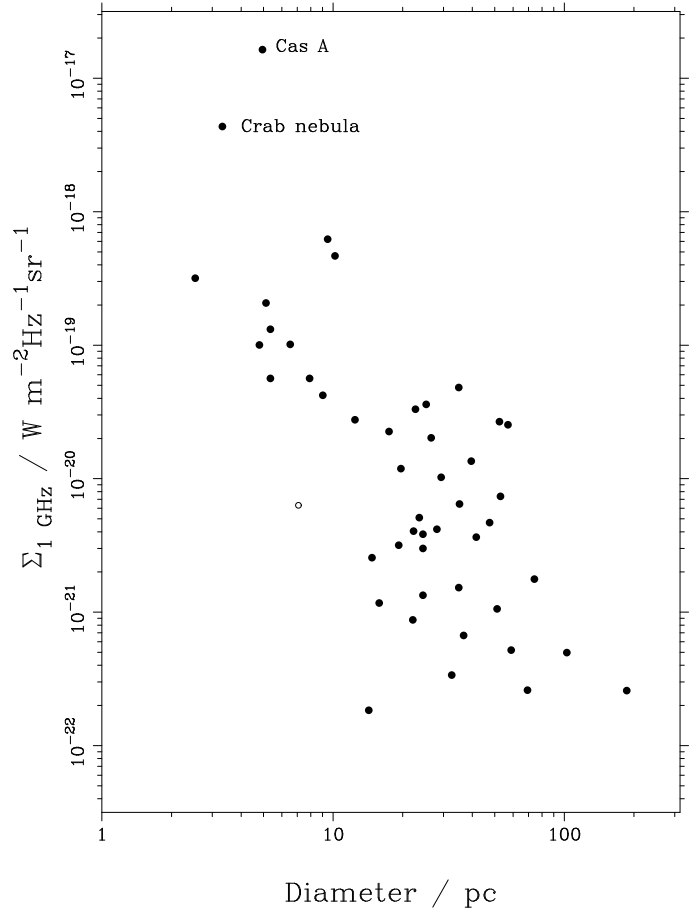}}
\caption{The surface brightness/diameter ($\itSigma{-}D$) relation for 47
Galactic SNRs with known distances (see Table~\ref{t:distances}), shown as
filled circles. The open circle shows the parameters of RX
J$0852{\cdot}0{-}4622$ ($=$\SNR(266.2)-(1.2)), if it is at a distance of 200~pc
(see text for discussion). Note that the lower left part of this diagram is
likely to be seriously affected by selection effects.\label{f:sigmad}}
\end{figure}

\begin{figure}
\centerline{\includegraphics[angle=270,width=9cm]{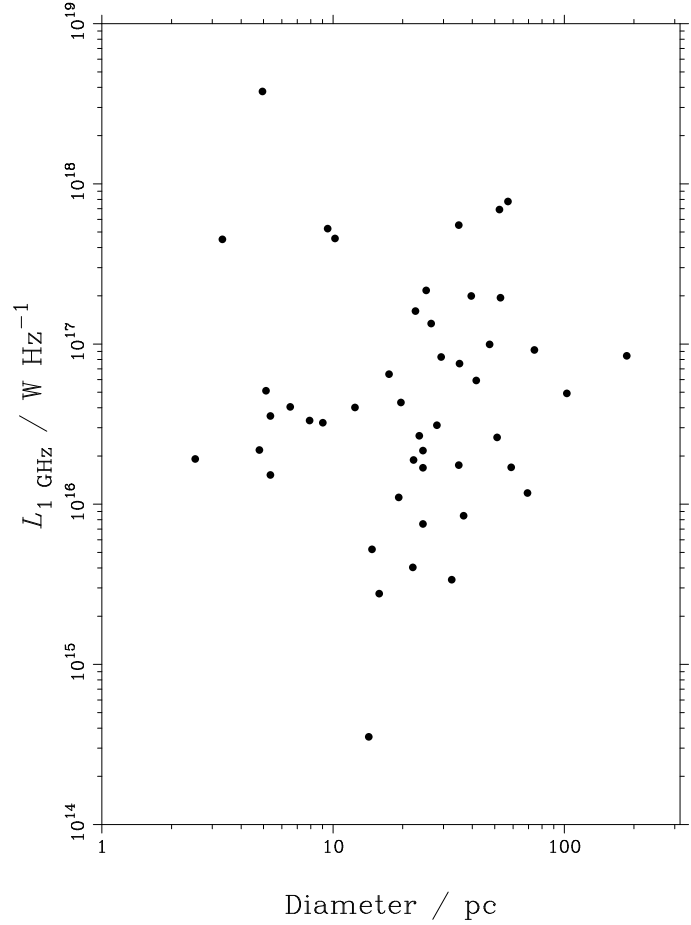}}
\caption{The luminosity/diameter ($L{-}D$) relation for 47 Galactic SNRs
with known distances (see Table~\ref{t:distances}).\label{f:ld}}
\end{figure}

\subsection{The $\itSigma{-}D$ and $L{-}D$ Relations}\label{s:sigmad}

Since distances are not available for all SNRs, many statistical studies of
Galactic SNRs have relied on the surface-brightness/diameter, or
`$\itSigma{-}D$' relation to derive distances for individual SNRs from their
observed flux densities and angular sizes. For remnants with known distances
($d$), and hence known diameters ($D$), physically large SNRs are fainter
(i.e.\ they have a lower surface brightness) than small remnants. Using this
correlation between $\itSigma$ and $D$ for remnants with known distances, a
physical diameter is deduced from the distance-independent {\em observed}
surface brightness of any remnant. Then a distance to the remnant can be
deduced from this diameter and the observed angular size of the remnant.

The $\itSigma{-}D$ relation for Galactic SNRs with known distances is shown in
Fig.~\ref{f:sigmad}. As discussed above (Section~\ref{s:distances}), the
distances to individual remnants are not homogeneous in quality, and many
depend on subjective interpretation of data.
Of the SNRs included in this figure, three are
`filled-centre' remnants (the Crab nebula ($=$\SNR(184.6)-(5.8)), 3C58
($=$\SNR(130.7)+(3.1)) and \SNR(74.9)+(1.2)). Nevertheless, Fig.~\ref{f:sigmad}
clearly shows a wide range of diameters for a given surface brightness, which
is a severe limitation in the usefulness of the $\itSigma{-}D$ relation for
deriving the diameters, and hence distances, to individual remnants. For a
particular surface brightness, the diameters of SNRs vary by up to about an
order of magnitude, or conversely, for a particular diameter, the range of
observed surface brightnesses seen varies by more than two orders of magnitude.

The correlation shown between surface brightness and diameter in
Fig.~\ref{f:sigmad} is, however, largely a consequence of the fact that it is a
plot of surface-brightness -- rather than luminosity -- against diameter, $D$.
Surface brightness is plotted, because it is the distance-independent
observable that is available for (almost) all SNRs, including those for which
distances are not available. For remnants whose distances are known, we can
instead consider the radio luminosity of the remnants. Since $\itSigma$ and
luminosity, $L$, depends on the flux density $S$, angular size $\theta$,
distance $d$ and diameter $D$, as
$$
  \itSigma \propto {S \over \theta^2} \quad{\rm and}\quad
  L \propto S d^2
$$
then
$$
   \itSigma \propto {L \over (\theta d)^2} \quad{\rm or}\quad
   \itSigma \propto {L \over D^2}.
$$
Thus, much of the correlation shown in the $\itSigma{-}D$ relation in
Fig.~\ref{f:sigmad} is due to the $D^{-2}$ bias that is inherent when plotting
$\itSigma$ against $D$, instead of $L$ against $D$. The $L{-}D$ relation for
Galactic SNRs with known distances in Fig.~\ref{f:ld} shows that there is wide
range of luminosities for SNRs of all diameters. Cas~A is the most luminous
Galactic SNR, but it appears to be at the edge of a wide distribution of
luminosities. The wide range of luminosities is perhaps not surprising, given
that the remnants are produced for a variety of types of supernovae, and that
they evolve in regions of ISM with a range of properties (e.g.\ density), which
may well effect the efficiency of the radio emission mechanism at work. For
example, some SNRs may initially evolve inside a low-density, wind-blown
cavity, and then collide with the much denser regions of the surrounding ISM.

Furthermore, the full range of intrinsic properties of SNRs may be even wider
than that shown in Figs~\ref{f:sigmad} and \ref{f:ld}, as the selection effects
discussed above mean that it is difficult to identify small and/or faint SNRs.
One specific example is the recently
identified SNR RX J$0852{\cdot}0{-}4622$ ($=$\SNR(266.2)-(1.2), see Aschenbach
1998), which may extend the range of properties of SNRs considerably (see
Duncan \& Green 2000). The surface brightness of RX J$0852{\cdot}0{-}4622$ at
1~GHz is $\approx 6 \times 10^{-22}$ {\sigmaunit}, which places it among the
faintest 20 per cent of catalogued remnants. If it is at a distance as small as
200~pc, as suggested by Aschenbach (1998) -- see also Redman et al.\ (2002) --
then its diameter would be only 7~pc, see Fig.~\ref{f:sigmad}. On the other
hand, the deficit of bright, large SNRs in Fig.~\ref{f:sigmad}
cannot be explained by any selection
effect, and so represents some real limit in the luminosity of remnants at a
particular diameter, related to the physics of the underlying radio emission
mechanism at work. This upper bound in the $\itSigma{-}D$ plane can be used to
derive an {\em upper limit} for the diameter of any SNR from its observed
surface brightness, and hence an upper limit on its distance. Another upper
bound on the distance to any remnant -- which may be as useful -- is to assume
it lies within the Galactic disc.

Case \& Bhattacharya (1998) derived a $\itSigma{-}D$ relation, based on the
distances available for 36 remnants -- not including filled-centre remnants --
from the 1996 version of the catalogue. They argue that
it is useful for deriving
distances for individual SNRs, with a fractional error of only 0.33, which is
very much smaller than the wide range in diameters for a given surface
brightness shown in Fig.~\ref{f:sigmad}. This optimistic result was only
obtained after excluding Cas~A from the remnants with known distances used to
derived the $\itSigma{-}D$ relation (as it was deemed to be sufficiently
different from other shell SNRs), and also after excluding 7 remnants with high
$z$-values (as three of these showed the largest deviation from the best-fit
$\itSigma{-}D$ relation). It is, however, difficult to decide {\em a priori}
whether a remnant is or is not to be included in the subset of remnants for
which Case \& Bhattacharya derived distances with relatively small
uncertainties.

Also, it is not clear that any best-fit $\itSigma{-}D$ relation -- not
withstanding selection effect problems -- actually represents the evolutionary
track of individual SNRs (see Berkhuijsen 1986). The distribution of SNRs with
known distances is a snapshot in time of a population of remnants, and
individual remnants may evolve in the $\itSigma{-}D$ plane in directions quite
different from the overall power law fitted to the overall distribution of SNRs
(or to the upper limit of the distribution). As a simple example, consider the
situation where SNRs have a range of intrinsic luminosities, expand with a
constant luminosity up to some particular diameter --  which varies for
different SNRs depending on their environment (e.g.\ the surrounding ISM
density, which influences their expansion speed, which may affect the
efficiency of radio emission mechanism) -- after which their radio luminosity
fades rapidly. In this case, the locus of the upper bound to the highest
surface brightness remnants for a particular diameter is related to where the
luminosities of different SNRs begin to decrease, and does not represent the
evolutionary track of any individual remnant. The current direction of the
evolutionary track of only one Galactic SNR, Cas~A, can be estimated from
available observations. The flux density of Cas~A is decreasing at
approximately 0.8\% year$^{-1}$ (Baars et al.\ 1977; Rees 1990), and its bulk
expansion is 0.22\% year$^{-1}$ (Ag\"ueros \& Green 1999; but see DeLaney
et al.\ 2004 for alternative expansion timescales). These observations suggest
Cas~A is following a track with $\Sigma \propto D^{-5.6}$, although the
uncertainty in this slope is large (nominally $\pm 1.2$, if the uncertainty in
the secular flux density decrease is taken as 0.2\% year$^{-1}$, and the expansion
timescale of the remnant is between 400 to 500 years).

\subsection{An Aside: Extragalactic Selection Effects}\label{s:extragalactic}

As noted above, a major problem with statistical studies of Galactic SNRs is
the difficulty of obtaining reliable distances for remnants. Studies of samples
of remnants in external galaxies are more straightforward in this respect, as
all the remnants are at a very similar distance. Given this, then it is
arguably more appropriate to consider the $L{-}D$ relation, rather than the
$\itSigma{-}D$, for extragalactic SNRs -- as the latter is not needed to
determine distances for individual remnants -- with the appropriate selection
effects. Then, for unresolved sources, in most radio studies the dominant
selection effect is a flux density limit, which corresponds to a fixed
luminosity limit for a particular galaxy.

In a recent study of the statistical properties of extragalactic SNRs in
several galaxies -- in addition to those in the Milky Way -- Arbutina et al.\
(2004) concluded that only in the case of M82 was there a good $L{-}D$
correlation, and hence a useful $\itSigma{-}D$ relation also. In other cases
there was a poor correlation between the luminosity and diameter of identified
SNRs (as noted in Section~\ref{s:sigmad} above for Galactic SNRs). However, in
their study Arbutina et al.\ have not correctly appreciated the observational
selection effects applicable to the sample of SNRs in M82. This sample of SNRs
was identified by Huang et al.\ (1994) from observations at 8.4~GHz. In Fig.~1
of Arbutina et al.\ a sensitivity limit corresponding to a luminosity of
$\approx 7 \times 10^{24}$ erg s$^{-1}$ Hz$^{-1}$ is plotted, which although
appropriate at 8.4~GHz, is not appropriate for the luminosities of the M82
sample of SNRs plotted in this figure, which are at 1~GHz. The true luminosity
limit from Huang et al.'s observations should be moved upwards on Arbutina et
al.'s Fig.~1 by $\approx 3$ (i.e.\ $\approx 8.4^{0.5}$ for a typical spectral
index of 0.5 to correct from 8.4 to 1~GHz). Moreover, the actual observational
selection effects provide a more stringent limit on the detectable luminosities
of the larger SNRs in the sample. Any remnants with a diameter of larger than
$\approx 3$~pc were resolved by Huang et al.'s observations, and a surface
brightness limit (equivalent to luminosity scaling as $D^2$), not a constant
luminosity limit is appropriate (see Fig.~2 of Huang et al., who show these
limits on a $\itSigma{-}D$ rather than a $L{-}D$ plot). Consequently, the
apparent range of luminosities of SNRs in M82 is strongly limited by selection
effects, particularly for the larger remnants. Indeed, it is noticeable that
the range of luminosities shown for the larger remnants in M82 appears smaller
than that of the smaller remnants, which seems unlikely to be real. Thus the
range of luminosities of SNRs in M82 is likely to extend to lower values, but
these objects have not been identified due to selection effects. In this case
the correlation between luminosity and diameter for remnants here is not
strong, as is the case in other galaxies, including our own. Consequently, the
$\itSigma{-}D$ relation for M82 is also affected by selection effects, and is
therefore of limited use.

\section{Galactic SNR Distribution}\label{s:distribution}

The distribution of SNRs in the Galaxy is of interest for many astrophysical
studies, particularly in relation to their energy input into the ISM and for
comparison with the distributions of possible progenitor populations. Such
studies are, however, not straightforward, due to observational selection
effects and the lack of reliable distance estimates available for most identified
remnants. In particular, all SNRs in the anti-centre (i.e.\ 2nd and 3rd
Galactic quadrants) are outside the Solar Circle, at large Galactocentric
radii, in regions where the background Galactic emission is low, so that low
surface brightness remnants are relatively easy to identify (see
Section~\ref{s:selection}). Without taking selection effects into account, the
larger number of fainter SNRs in the anti-centre leads to an apparently broad
distribution of Galactic SNRs in Galactocentric radius (e.g.\ the very broad
distribution of SNRs derived by Li et al.\ (1991), who included all SNRs in
their analyses). A more complicated method to derive the radial distribution of
Galactic SNRs is that used by Case \& Bhattacharya (1996, 1998), following a
method used by Narayan (1987) for pulsars. This relies on (i) assuming
catalogues of Galactic SNRs are complete for SNRs within a distance of 3~kpc;
(ii) using $\itSigma{-}D$ derived distances for the SNRs, and (iii) attempts to
correct for observational selection effects using a scaling factor that varies
in many bins across the disc of the Galaxy. However, this is difficult given
the uncertainties in the usefulness of $\itSigma{-}D$ relation discussed above,
and the necessity of deconvolving selection effects from the observed
distribution of SNRs. An alternative approach is to investigate the
distribution of SNRs in Galactic coordinates, restricting the studies to
relatively bright remnants, for which current catalogues are thought to be
complete. van den Bergh (1988a,b) discussed the distribution of observed SNRs
and noted that high surface brightness remnants (in this case taken to be
$\itSigma_{\rm 1~GHz} > 3 \times 10^{-21}$ \sigmaunit) are concentrated in a
thin nuclear disc when plotted in Galactic coordinates. As noted by F\"urst's
comments to van den Bergh (1988b), this conclusion is strengthened by a more
realistic surface brightness completeness limit.

\begin{figure}
\centerline{\includegraphics[width=8cm,clip=]{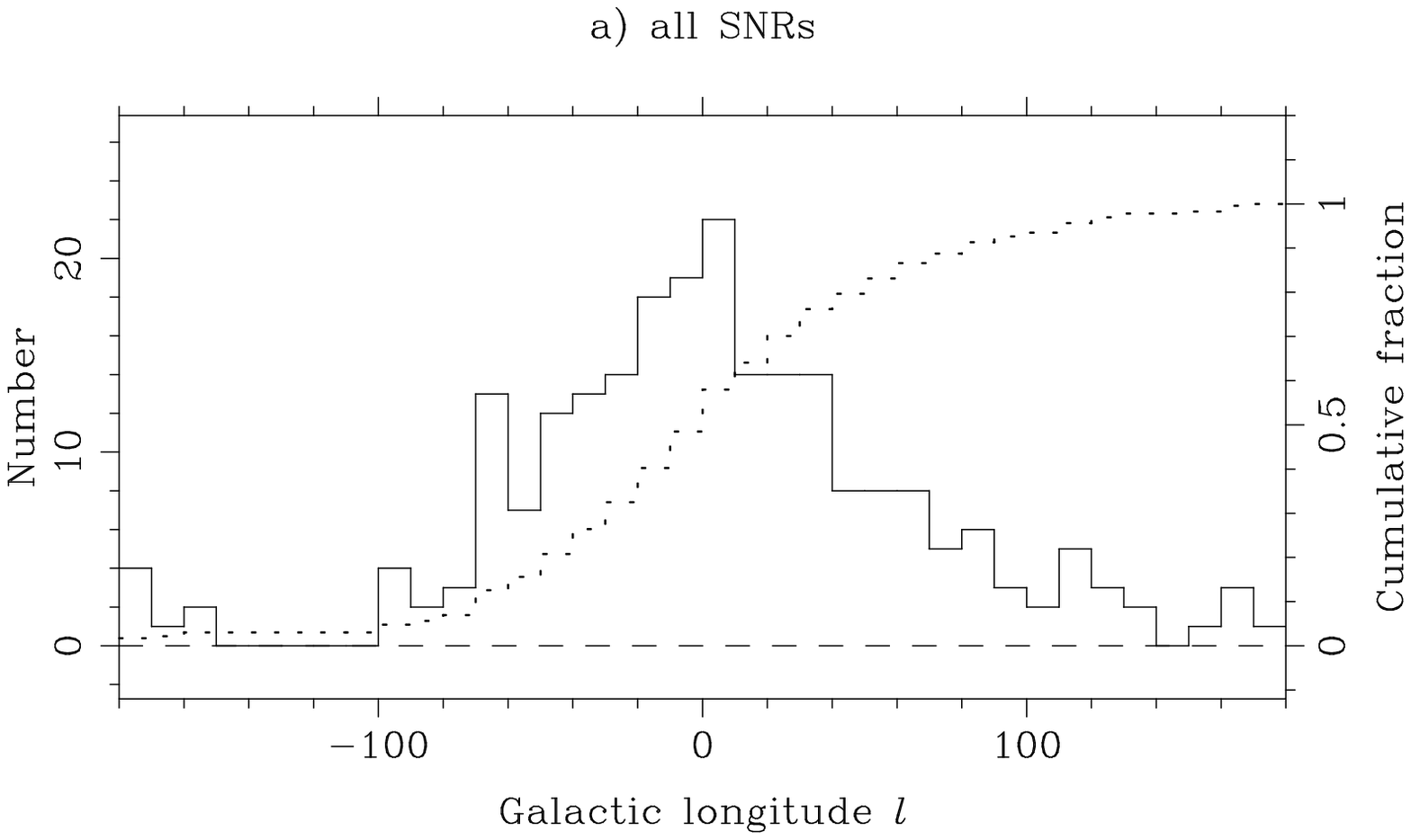}}
  \quad\\
\centerline{\includegraphics[width=8cm,clip=]{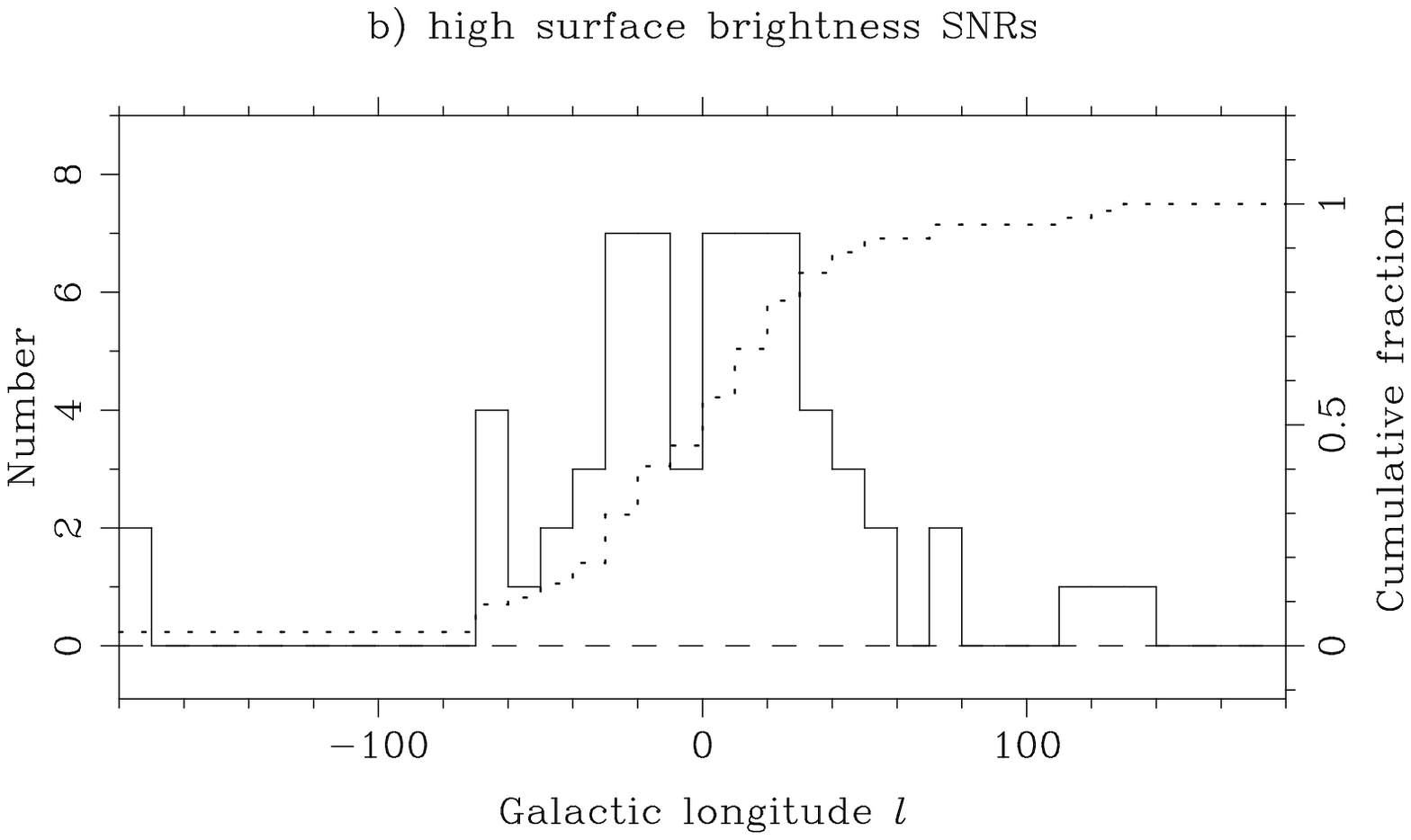}}
\caption{The distribution in Galactic longitude of (top) all 231 Galactic SNRs,
and (bottom) the 64 high surface brightness SNRs with $\itSigma_{\rm 1~GHz} >
10^{-20}$ {\sigmaunit}. Each plot shows as a solid line a histogram of the
observed longitudes of the remnants (left scale), and as a dotted line for
cumulative fraction (right scale).\label{f:all-bright}}
\end{figure}

\begin{figure}
\centerline{\includegraphics[width=8cm,clip=]{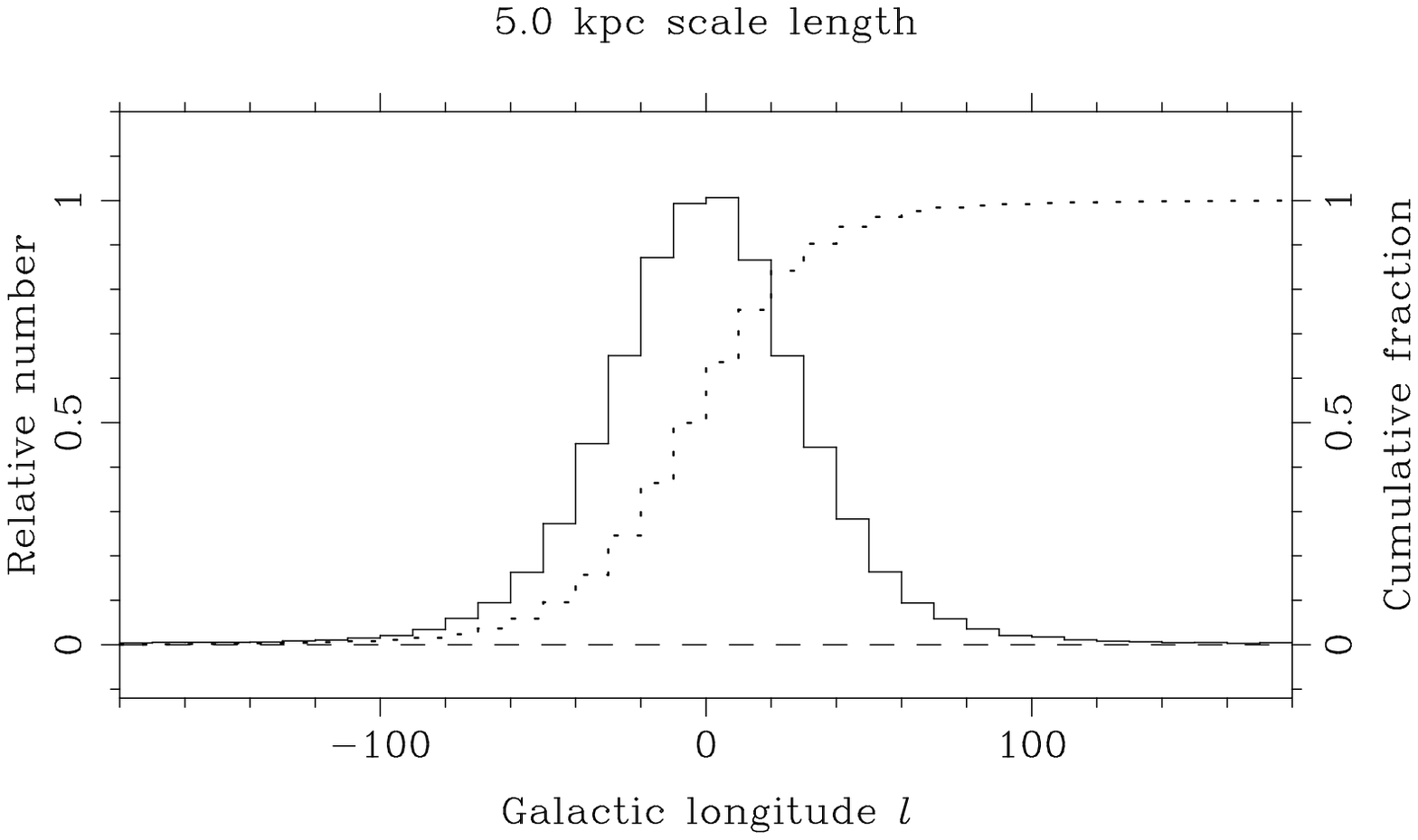}}
\quad\\
\centerline{\includegraphics[width=8cm,clip=]{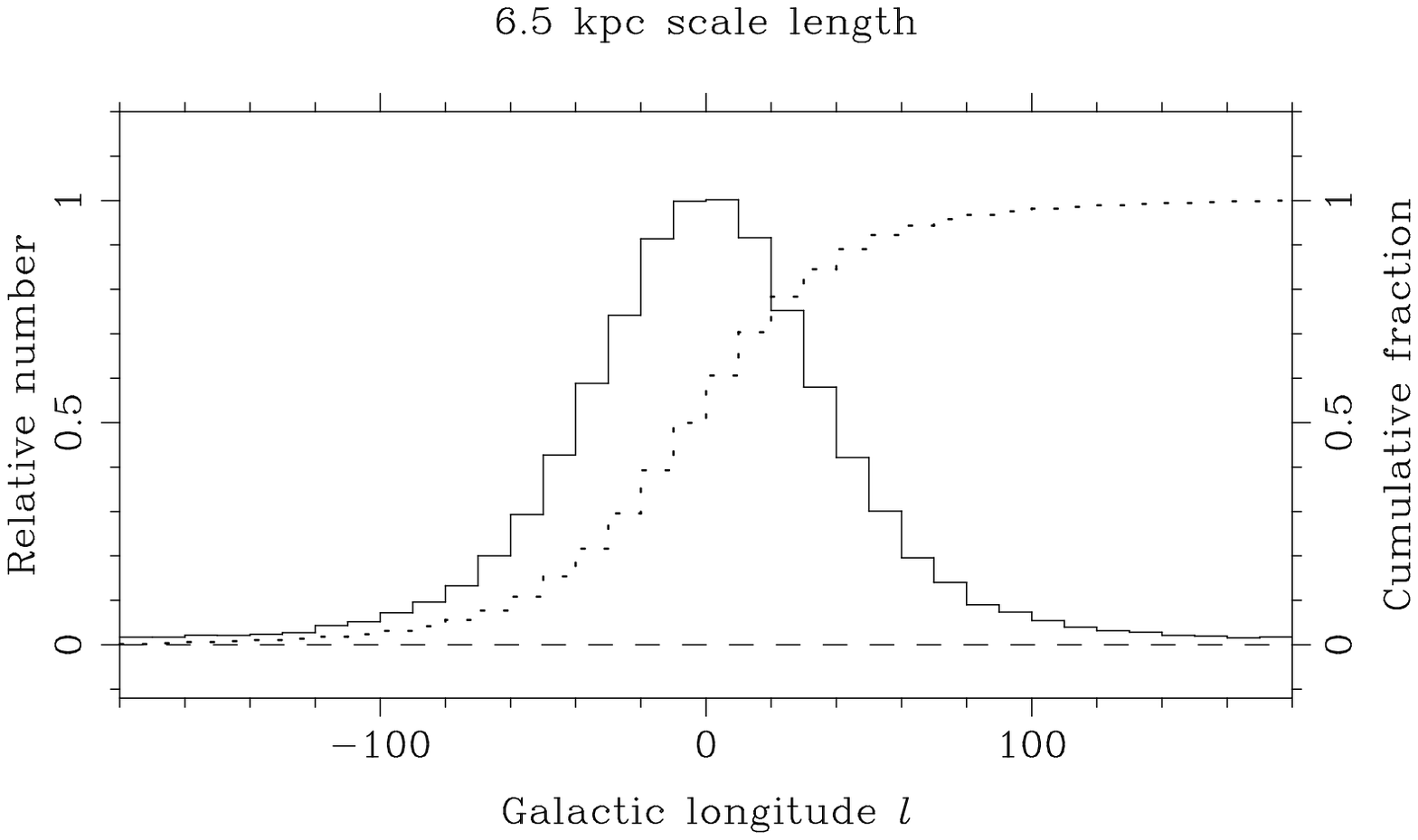}}
\quad\\
\centerline{\includegraphics[width=8cm,clip=]{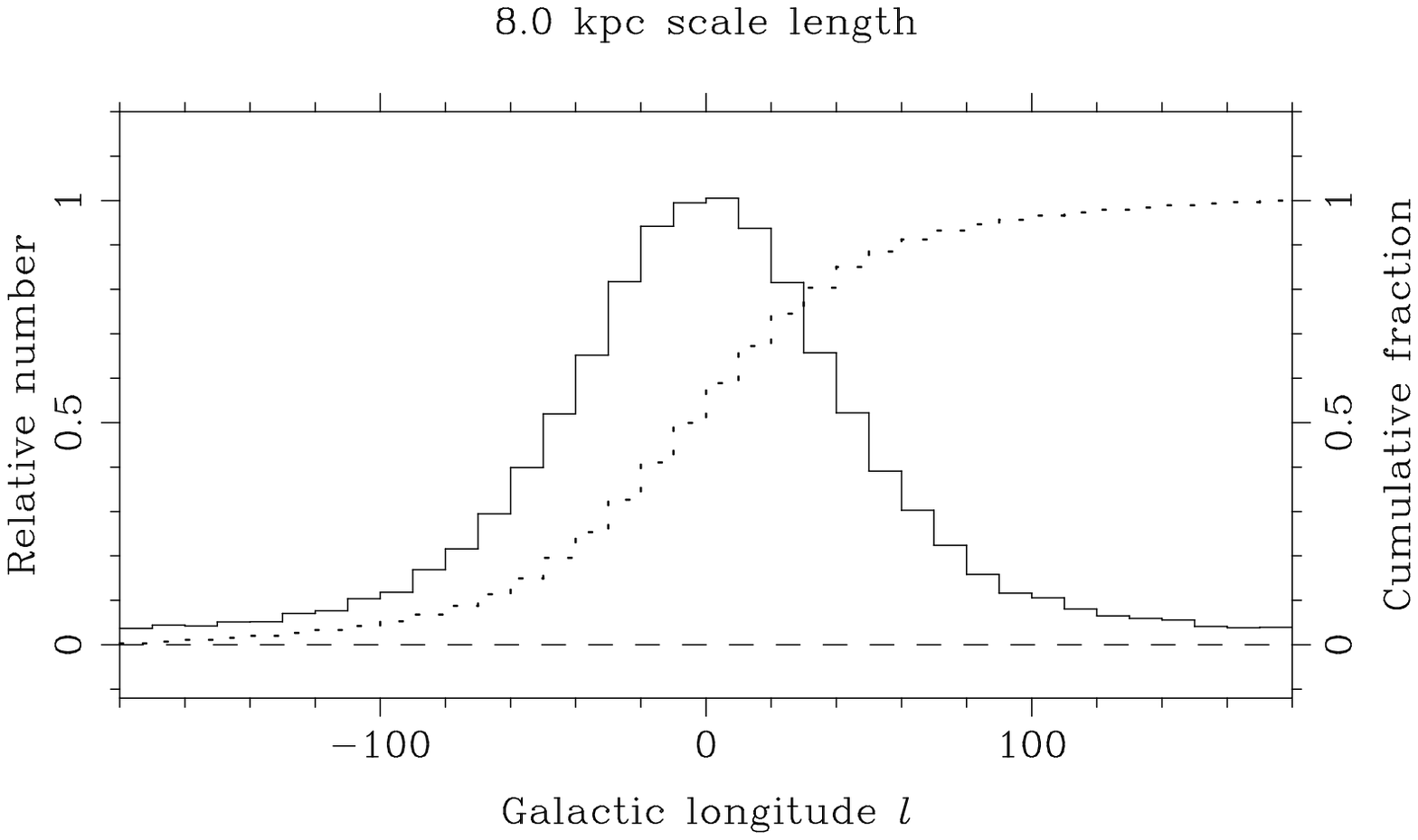}}
\caption{Model distribution in Galactic longitude of Gaussian distributions of
SNRs with three different Galactocentric radius scale lengths (cf.\ the
observed distribution in Fig.~\ref{f:all-bright}).\label{f:model}}
\end{figure}

More quantitatively -- following the method of Li et al., but using an
appropriate selection brightness cut-off -- the observed distribution of bright
SNRs with Galactic longitude can be compared with that expected from various
models. A major advantage of this method is that it avoids the problem that we
lack accurate distances to individual SNRs, although on the other hand it uses
only a sub-set of the known Galactic SNRs. Fig.~\ref{f:all-bright} shows the
observed distributions with Galactic longitude of all Galactic SNRs, and of the
64 remnants which have $\itSigma_{\rm 1~GHz} > 10^{-20}$ {\sigmaunit}  (this is
a similar number to the 71 remnants used in a similar study presented in Green
1996a, which used a slightly lower surface brightness cut-off applied to the
SNR catalogue of Green 1996b). By applying the surface brightness cut-off, so
that the surface brightness selection effect is not important, it is clear that
the distribution of Galactic remnants is actually much more concentrated
towards $l=0^\circ$ than if all remnants are considered (cf.\ Fig.~\ref{f:lb}).
Fig.~\ref{f:all-bright} shows evidence for a deficit of SNRs near
$l=350^\circ$, which may be a true deficit if there is a decrease in the space
density of SN progenitors towards the Galactic centre. However, it may also be,
in part at least, due to the difficulty of finding remnants in this region of
the Galactic plane, due to the very complex background emission and confusing
Galactic sources (e.g.\ {\HII} regions). Any remaining incompleteness in
current catalogues, both for the surface brightness and angular diameter
selection effects, are expected to be worse closer to $b=0^\circ$ (because of
the increased confusion in the case of the surface brightness selection effect,
and the longer line-of-sight through the Galaxy for missing small, i.e.\ young
but distant remnants). Thus, the true distribution in $l$ is likely to be
somewhat {\sl narrower} than is indicated in Fig.~\ref{f:all-bright}. For
comparison with the observed distributions in Galactic longitude, simple Monte
Carlo models of the distribution of SNRs in the disc of the Galaxy were
constructed assuming a simple, circularly symmetric, Gaussian distribution,
where the probability distribution varies with Galactocentric radius, $R$, as
$$
  \propto {\rm e}^{-(R/\sigma)^2},
$$
(where $\sigma$ is the Gaussian Galactocentric scale length, assuming the
distance to the Galactic Centre is 8.5~kpc). Fig.~\ref{f:model} shows plots of
the expected distribution of SNRs in Galactic longitude of three such models
for different scale lengths. As noted above, the true distribution is likely to
be somewhat narrower than that derived from the observations, due to residual
selection effects, so that this scale length is an upper limit. A $\chi^2$
comparison of the observed and model cumulative distributions indicates that
for this simple model, a scale length of $\approx 6.5$~kpc best matches the
observed distribution of high brightness SNRs.

The model distribution of SNRs derived above should, however, be interpreted
cautiously, as not only is it a simplistic model without spiral arm structures,
but also it is a model of the distribution of {\em observed} remnants. It is
far from clear what factors affect the brightness and lifetime of appreciable
radio emission from SNRs -- i.e.\ their observability at radio wavelengths --
and hence how close the distribution of observed SNRs is to the parent
supernovae distribution. The distribution of SNRs could reflect the
distribution of, for example, the density of the ISM, or the Galactic magnetic
field, if these are important factors in determining the brightness and lifetime
of radio emission from SNRs.

\section{Conclusions}

Here I have presented a recent catalogue of 231 Galactic SNRs, and have
discussed the selection effects that apply to the identification of remnants.
Both surface brightness and angular size selection effects are important, and
these need to be borne in mind when statistical studies of Galactic SNRs are
made. One consequence of the current angular size selection effect is that few
young but distant remnants have yet been identified in the Galaxy. These
objects are likely to be in complex regions of the Galactic plane, and further
observations -- using a wide range of radio wavelengths and/or the combination
of radio and other wavelengths --  are required to identify these missing
objects. For remnants with known distances, the intrinsic range of luminosity
of Galactic SNRs is large, which combined with selection effects, means that
the $\itSigma{-}D$ relation is of limited use for determining distances to
individual remnants, or for statistical studies.

\section*{Acknowledgements}

I am grateful to many colleagues for numerous comments on, and corrections to,
the various versions of the Galactic SNR catalogue, and for comments on early
versions of this work. I also thank the referee, Chris Salter, for his useful
and detailed report. This paper was finalised during an extended visit to the
National Radio Astronomy Observatory, Socorro, NM, USA in 2004 September, and I
am very grateful for their hospitality at that time. This research has made use
of NASA's Astrophysics Data System Bibliographic Services.


\appendix
\section{The Galactic SNR catalogue: 2004 January version}\label{s:appendix}

This appendix presents a catalogue of Galactic supernova remnants. This
catalogue is an updated version of those presented in detail in Green (1984,
1988) and in summary form in Green (1991, 1996b) and Stephenson \& Green
(2002). Detailed versions of this catalogue have been made available on the
World-Wide-Web since 1993 November (with subsequent versions of 1995 July, 1996
August, 1998 September, 2000 August, 2001 December and 2004 January). This, the
2004 January version of the catalogue, contains 231 SNRs. The detailed version
of this catalogue is available at
\\ \centerline{\tt http://www.mrao.cam.ac.uk/surveys/snrs/}\\ \\
which contains over a thousand references for the individual SNRs.

For each remnant in the catalogue the following parameters are given.
\begin{itemize}
\item {\bf Galactic Coordinates} of the source centroid, quoted to the nearest
tenth of a degree as is conventional. (Note: in this catalogue additional
leading zeros are not used.)
\item {\bf Right Ascension} and {\bf Declination} of the source centroid. The
accuracy of the quoted values depends on the size of the remnant, for small
remnants they are to the nearest few seconds of time and the nearest minute of
arc respectively, whereas for larger remnants they are rounded to coarser
values, but are in every case sufficient to specify a point within the boundary
of the remnant. These coordinates are almost always deduced from radio maps
rather than from X-ray or optical observations, and are for J2000.0.
\item {\bf Angular Size} of the remnant, in arcminutes, usually taken from the
highest resolution radio map available. The boundary of most remnants
approximates reasonably well to a circle or an ellipse, a single value is
quoted for the angular size of the more nearly circular remnants, which is the
diameter of a circle with an area equal to that of the remnant, but for
elongated remnants the product of two values is quoted, and these are the major
and minor axes of the remnant boundary modelled as an ellipse. In a few cases
an ellipse is not a satisfactory description of the boundary of the object
(refer to the description of the individual object given in its catalogue
entry), although an angular size is still quoted for information. For
`filled-centre' remnants the size quoted is for the largest extent of the
observed radio emission, not, as at times has been used by others, the
half-width of the centrally brightened peak.
\item {\bf Type} of the SNR: `S' or `F' if the remnant shows a `shell' or
`filled-centre' structure, or `C' if it shows `composite' (or `combination')
radio structure with a combination of shell and filled-centre characteristics;
or `S?', `F?' or `C?', respectively, if there is some uncertainty, or `?' in
several cases where an object is conventionally regarded as an SNR even though
its nature is poorly known or not well understood. (Note: the term `composite'
has been used in a different sense by some authors, to describe SNRs with shell
radio and centrally-brightened X-ray morphologies. An alternative term used to
describe such remnants is `mixed morphology', see Rho \& Petre 1998.)
\item {\bf Flux Density} of the remnant at 1~GHz in jansky. This is {\sl not}
a measured value, but is deduced from the observed radio frequency spectrum of
the source. The frequency of 1~GHz is chosen because flux density measurements
at frequencies both above and below this value are usually available.
\item {\bf Spectral Index} of the integrated radio emission from the remnant,
$\alpha$ (here defined in the sense, $S \propto \nu^{-\alpha}$, where $S$ is
the flux density at a frequency $\nu$), either a value that is quoted in the
literature, or one deduced from the available integrated flux densities of the
remnant. For several SNRs a simple power law is not adequate to describe their
radio spectra, either because there is evidence that the integrated spectrum is
curved or the spectral index varies across the face of the remnant. In these
cases the spectral index is given as `varies' (refer to the description of the
remnant and appropriate references in the detailed catalogue entry for more
information). In some cases, for example where the remnant is highly confused
with thermal emission, the spectral index is given as `?' since no value can be
deduced with any confidence.
\item {\bf Other Names} that are commonly used for the remnant. These are
given in parentheses if the remnant is only a part of the source. For some
remnants, notably the Crab nebula, not all common names are given.
\end{itemize}
A summary of the data available for all 231 remnants in the catalogue is given
in Table~A1.

In the detailed listings, available on the World-Wide-Web, notes on a variety
of topics are given for each remnant.  First, it is noted if other Galactic
coordinates have at times been used to label it (usually before good
observations have revealed the full extent of the object), if the SNR is
thought to be the remnant of a historical SN, or if the nature of the source as
an SNR has been questioned (in which case an appropriate reference is usually
given later in the entry). Brief descriptions of the remnant from the available
radio, optical and X-ray observations as applicable are then given, together
with notes on available distance determinations, and any point sources or
pulsars in or near the object (although they may not necessarily be related to
the remnant). Finally, appropriate references to observations are given for
each remnant, complete with journal, volume, page, and a short description of
what information each paper contains (for radio observations these include the
telescopes used, the observing frequencies and resolutions, together with any
flux density determinations). These references are {\sl not} complete, but
cover representative and recent observations of the remnant -- up to the end of
2003 -- and they should themselves include references to earlier work. The
references do not generally include large observational surveys -- of
particular interest in this respect are: the Effelsberg 100-m survey at 2.7~GHz
of the Galactic plane $358^\circ < l <  240^\circ$, $|b| < 5^\circ$ by Reich et
al.\ (1990) and F\"urst et al.\ (1990), reviews of the radio spectra of some
SNRs by Kassim (1989), Kovalenko, Pynzar' \& Udal'tsov (1994) and Trushkin
(1998), the Parkes 64-m survey at 2.4~GHz of the Galactic plane $238^\circ < l
< 365^\circ$, $|b| < 5^\circ$ by Duncan et al.\ (1995) and Duncan et al.\
(1997), the Molonglo Galactic plane survey at 843~MHz of $245^\circ < l <
355^\circ$, $|b| < 1\fdg5$ by Green et al.\ (1999), the survey of $345^\circ <
l < 255^\circ$, $|b|<5^\circ$ at 8.35 and 14.35~GHz by Langston et al.\ (2000),
reviews of {\sl Einstein} X-ray imaging and spectroscopic observations of
Galactic SNRs by Seward (1990) and Lum et al.\ (1992) respectively, surveys of
{\sl IRAS} observations of SNRs and their immediate surroundings by Arendt
(1989) and by Saken, Fesen \& Shull (1992), the survey of {\HI} emission
towards SNRs by Koo \& Heiles (1991), and the catalogue by Fesen \& Hurford
(1996) of UV/optical/infra-red lines identified in SNRs. The detailed version
of the catalogue also including notes on the objects no longer thought to be
SNRs, and on many possible and probable remnants in the literature. It should
also be noted that: (i) some radio continuum and {\HI} loops in the Galactic
plane (e.g.\ Berkhuijsen 1973) may be parts of very large, old SNRs, but they
have not been included in the catalogue (see also Combi et al.\ 1995;
Maciejewski et al.\ 1996; Kim \& Koo 2000; Normandeau et al.\ 2000; Woermann,
Gaylard \& Otrupcek 2001; Stil \& Irwin 2001; Uyan{\i}ker \& Kothes 2002), (ii)
the distinction between filled-centre remnants and pulsar wind nebula is not
clear, and isolated, generally faint, pulsar wind nebulae are also not included
in the catalogue (e.g.\ Gaensler et al.\ 1998b; Giacani et al.\ 2001; Jones,
Stappers \& Gaensler 2002; Braje et al.\ 2002; Gaensler et al.\ 2003).

\clearpage
\newcount\linesdone
\global\linesdone=0
\newcount\processed
\global\processed=0
%
%
\def\captiontext{Galactic Supernova Remnants: summary data.}
\def\tops{%
  \setbox0=\vbox\bgroup%
  \centerline{\small{\bf Table~A1}. \captiontext}
  \centerline{\hrulefill}
  \smallskip
  \centerline{%
    \hbox to 0.06\hsize{\hfil$l$\enskip}%
    \hbox to 0.065\hsize{\hfil$b$\enskip}%
    \hbox to 0.19\hsize{\hss\quad RA (J2000) Dec\hss}%
    \hbox to 0.12\hsize{\hfil size\hfil}%
    \hbox to 0.05\hsize{type\hfil}%
    \hbox to 0.11\hsize{\hfil Flux at\hfil}%
    \hbox to 0.10\hsize{\hfil spectral\hfil}%
    \hbox to 0.325\hsize{\enspace other \hfil}%
    \hfill
  }
  \centerline{%
    \hbox to 0.125\hsize{\hfil}%
    \hbox to 0.11\hsize{\hfil$({\rm h}$\enskip${\rm m}$\enskip${\rm s})$}%
    \hbox to 0.08\hsize{\hfil$({}^\circ$\quad${}'$)}%
    \hbox to 0.12\hsize{\hfil /arcmin\hfil}%
    \hbox to 0.05\hsize{}%
    \hbox to 0.11\hsize{\hfil 1~GHz/Jy\hfil}%
    \hbox to 0.10\hsize{\hfil index\hfil}%
    \hbox to 0.325\hsize{\enspace name(s)\hfil}%
    \hfill
  }
  \centerline{\hrulefill} %
  \medskip
  \egroup\box0
}
%
%
%
%
%
%
%
%
%
%
%
%
\def\LONGITUDE #1 {\def\ldegrees{#1}}
\def\LATITUDE #1 {\def\bdegrees{#1}}
\def\RAHMS #1 #2 #3 {\def\rahms{#1~#2~#3}}
\def\DECDM #1 #2 {\def\decdm{#1~#2}}
\def\SIZE #1 {\def\size{#1}}
\def\ALPHA #1 {\def\spectralindex{#1}}
\def\FLUX1GHZ #1 {\def\flux1GHz{#1}}
\def\TYPE #1 {\def\type{#1}}
\def\NAMES #1\par{\def\names{#1}\ifnum\linesdone=0\tops\fi%
  \global\advance\linesdone by 1
  \global\advance\processed by 1
  \centerline{%
    \hbox to 0.06\hsize{\hfil$\ldegrees$}%
    \hbox to 0.065\hsize{\hfil$\bdegrees$}%
    \hbox to 0.11\hsize{\hfil$\rahms$}%
    \hbox to 0.08\hsize{\hfil$\decdm$}%
    \hbox to 0.12\hsize{\hfil$\size$\hfil}%
    \hbox to 0.05\hsize{\enspace\type\hfil}%
    \hbox to 0.11\hsize{\hfil\flux1GHz\hfil}%
    \hbox to 0.10\hsize{\quad\spectralindex\hfil}%
    \hbox to 0.325\hsize{\enspace\names\hfil}%
    \hfill
  }
  \setbox0=\vbox\bgroup \eightpoint\hfuzz=20pt
}
%
%
\def\NOTES{\par}
\def\RADIO{\par}
\def\XRAY{\par}
\def\OPTICAL{\par}
\def\DISTANCE{\par}
\def\POINT{\par}
\def\REFERENCE{\par}
%
%
\def\PM/{\pm }
\def\X/{\times }
\def\I/{{\small I}}
\def\II/{{\small II}}
\def\III/{{\small III}}
\def\V/{{\small V}}
\def\VV/{{\small X}}       
\def\XIV/{{\small XIV}}
\def\AD/{{\small AD}}
\def\Halpha/{{H$\alpha$}}
\def\HCOplus/{{HCO$^+$}}
\def\etal/{{et al.}}
\let\eightpoint=\footnotesize
%
%
\def\DATE #1\par{
  \egroup
  \ifnum\linesdone=35
    \global\linesdone=0
    \global\processed=0
    \centerline{\hrulefill}
    \vfill\eject
    \def\captiontext{(continued).}
  \fi
  \ifnum\processed=5
    \vskip 6pt plus 2pt minus 1pt
    \global\processed=0
  \fi
}
%
%
\LONGITUDE 0.0 \LATITUDE +0.0
\RAHMS 17 45 44 \DECDM -29 00
\SIZE 3.5\X/2.5 \TYPE S
\FLUX1GHZ 100? \ALPHA 0.8?
\NAMES Sgr A East

\RADIO Non-thermal shell, in complex region, interacting with molecular material to the west.
\DATE 23 Jan 2004

\LONGITUDE 0.3 \LATITUDE +0.0
\RAHMS 17 46 15 \DECDM -28 38
\SIZE 15\X/8 \TYPE S
\FLUX1GHZ 22 \ALPHA 0.6
\NAMES

\NOTES Has been called G0.33$+$0.04 and G0.4$+$0.1.
\RADIO Bilateral shell, near Galactic Centre.
  observations.
\DATE 1 Aug 2000

\LONGITUDE 0.9 \LATITUDE +0.1
\RAHMS 17 47 21 \DECDM -28 09
\SIZE 8 \TYPE C
\FLUX1GHZ 18? \ALPHA varies
\NAMES

\RADIO Flat spectrum core within steep spectrum shell.
\XRAY Central core, with non-thermal spectrum.
\DATE 23 Jan 2004

\LONGITUDE 1.0 \LATITUDE -0.1
\RAHMS 17 48 30 \DECDM -28 09
\SIZE 8 \TYPE S
\FLUX1GHZ 15 \ALPHA 0.6?
\NAMES

\NOTES Has been called G1.05$-$0.1 and G1.05$-$0.15.
\RADIO Incomplete shell, to the S of Sgr~D.
\XRAY Possibly detected.
  masers observations.
\DATE 17 Oct 2001

\LONGITUDE 1.4 \LATITUDE -0.1
\RAHMS 17 49 39 \DECDM -27 46
\SIZE 10 \TYPE S
\FLUX1GHZ 2? \ALPHA ?
\NAMES

\RADIO Shell, brightest in E.
\DATE 11 Jan 2004

\LONGITUDE 1.9 \LATITUDE +0.3
\RAHMS 17 48 45 \DECDM -27 10
\SIZE 1.2 \TYPE S
\FLUX1GHZ 0.6 \ALPHA 0.7
\NAMES

\RADIO Shell, brighter to the N.
\DATE 13 Aug 1998

\LONGITUDE 3.7 \LATITUDE -0.2
\RAHMS 17 55 26 \DECDM -25 50
\SIZE 14\X/11 \TYPE S
\FLUX1GHZ 2.3 \ALPHA 0.65
\NAMES

\NOTES Hase been called G003.8$-$00.3.
\RADIO Double arc.
\DATE 13 Aug 1998

\LONGITUDE 3.8 \LATITUDE +0.3
\RAHMS 17 52 55 \DECDM -25 28
\SIZE 18 \TYPE S?
\FLUX1GHZ 3? \ALPHA 0.6
\NAMES

\RADIO Incomplete shell.
\DATE 11 Jan 2004

\LONGITUDE 4.2 \LATITUDE -3.5
\RAHMS 18 08 55 \DECDM -27 03
\SIZE 28 \TYPE S
\FLUX1GHZ 3.2? \ALPHA 0.6?
\NAMES

\RADIO Elongated shell.
\DATE 19th May 1992

\LONGITUDE 4.5 \LATITUDE +6.8
\RAHMS 17 30 42 \DECDM -21 29
\SIZE 3 \TYPE S
\FLUX1GHZ 19 \ALPHA 0.64
\NAMES Kepler, SN1604, 3C358

\NOTES This is the remnant of Kepler's SN of \AD/1604.
\RADIO Incomplete shell, brighter to the N.
\OPTICAL Faint filaments.
\XRAY Shell, brighter to the N.
\DISTANCE Optical expansion and proper motion indicates about 2.9~kpc, H\I/ observations suggest 3.4 to 6.4~kpc.
\DATE 11 Jan 2004

\LONGITUDE 4.8 \LATITUDE +6.2
\RAHMS 17 33 25 \DECDM -21 34
\SIZE 18 \TYPE S
\FLUX1GHZ 3 \ALPHA 0.6
\NAMES

\NOTES Has been called G4.5$+$6.2.
\RADIO Faint shell.
\DATE 12 Oct 2001

\LONGITUDE 5.2 \LATITUDE -2.6
\RAHMS 18 07 30 \DECDM -25 45
\SIZE 18 \TYPE S
\FLUX1GHZ 2.6? \ALPHA 0.6?
\NAMES

\RADIO Poorly resolved shell.
\DATE 19th May 1992

\LONGITUDE 5.4 \LATITUDE -1.2
\RAHMS 18 02 10 \DECDM -24 54
\SIZE 35 \TYPE C?
\FLUX1GHZ 35? \ALPHA 0.2?
\NAMES Milne 56

\NOTES Part been called G5.3$-$1.0. Has been suggested that this is not a SNR.
\RADIO Incomplete shell, including wide `v' of emission to east with small flat-spectrum source at apex.
\OPTICAL Detected.
\XRAY Pulsar detected, with faint extension.
\DISTANCE H\I/ absorption suggests $>4.3$~kpc.
\POINT Pulsar associated with flat spectrum source.
  including polarization.
\DATE 9 Dec 2001

\LONGITUDE 5.9 \LATITUDE +3.1
\RAHMS 17 47 20 \DECDM -22 16
\SIZE 20 \TYPE S
\FLUX1GHZ 3.3? \ALPHA 0.4?
\NAMES

\RADIO Asymmetric shell.
\DATE 19th May 1992

\LONGITUDE 6.1 \LATITUDE +1.2
\RAHMS 17 54 55 \DECDM -23 05
\SIZE 30\X/26 \TYPE F
\FLUX1GHZ 4.0? \ALPHA 0.3?
\NAMES

\NOTES Has been called G6.1$+$1.15.
\RADIO Faint, diffuse emission.
\DATE 28 Jun 1995

\LONGITUDE 6.4 \LATITUDE -0.1
\RAHMS 18 00 30 \DECDM -23 26
\SIZE 48 \TYPE C
\FLUX1GHZ 310 \ALPHA varies
\NAMES W28

\NOTES Has been called G6.6$-$0.2.
\RADIO Several non-thermal sources in a ring, with flat spectrum core.
\OPTICAL Filaments.
\XRAY Diffuse emission from most of the remnant.
\POINT Young pulsar near edge of remnant, but not thought to be related.
\DISTANCE H\I/ observations suggest 1.9~kpc.
  plus Einstein image of central region.
  and comparison with other observations.
\DATE 11 Jan 2004

\LONGITUDE 6.4 \LATITUDE +4.0
\RAHMS 17 45 10 \DECDM -21 22
\SIZE 31 \TYPE S
\FLUX1GHZ 1.3? \ALPHA 0.4?
\NAMES

\RADIO Faint asymmetric shell.
\DATE 19th May 1992

\LONGITUDE 7.0 \LATITUDE -0.1
\RAHMS 18 01 50 \DECDM -22 54
\SIZE 15 \TYPE S
\FLUX1GHZ 2.5? \ALPHA 0.5?
\NAMES

\NOTES Has been called G7.06$-$0.12.
\RADIO Double rim, brightest in W, confused by bright H\II/ region M20 in SE.
\DATE 10th October 2001

\LONGITUDE 7.7 \LATITUDE -3.7
\RAHMS 18 17 25 \DECDM -24 04
\SIZE 22 \TYPE S
\FLUX1GHZ 11 \ALPHA 0.32
\NAMES 1814$-$24

\RADIO Shell, with high polarization.
  with polarization, plus review of flux densities.
\DATE 20 Aug 1996

\LONGITUDE 8.7 \LATITUDE -5.0
\RAHMS 18 24 10 \DECDM -23 48
\SIZE 26 \TYPE S
\FLUX1GHZ 4.4 \ALPHA 0.3
\NAMES

\RADIO Asymmetric shell.
\DATE 19th May 1992

\LONGITUDE 8.7 \LATITUDE -0.1
\RAHMS 18 05 30 \DECDM -21 26
\SIZE 45 \TYPE S?
\FLUX1GHZ 80 \ALPHA 0.5
\NAMES (W30)

\RADIO Clumpy non-thermal shell, with low-frequency turnover.
\XRAY Northern edge detected.
\POINT Pulsar inside western edge.
  plus review of flux densities.
\DATE 27 Jun 1995

\LONGITUDE 9.8 \LATITUDE +0.6
\RAHMS 18 05 08 \DECDM -20 14
\SIZE 12 \TYPE S
\FLUX1GHZ 3.9 \ALPHA 0.5
\NAMES

\RADIO Asymmetric shell.
\DATE 13 Aug 1998

\LONGITUDE 11.2 \LATITUDE -0.3
\RAHMS 18 11 27 \DECDM -19 25
\SIZE 4 \TYPE C
\FLUX1GHZ 22 \ALPHA 0.6
\NAMES

\NOTES Probably associated with the SN of \AD/386.
\RADIO Symmetrical clumpy shell, with flatter spectrum core.
\XRAY Shell, with hard spectrum centrally brightened region around pulsar.
\POINT Central pulsar.
\DISTANCE H\I/ absorption indicates 5~kpc.
\DATE 11 Jan 2004

\LONGITUDE 11.4 \LATITUDE -0.1
\RAHMS 18 10 47 \DECDM -19 05
\SIZE 8 \TYPE S?
\FLUX1GHZ 6 \ALPHA 0.5
\NAMES

\RADIO Incomplete shell, possibly with central core.
  and Parkes 64-m at 5~GHz ($4': S=2.8$~Jy).
\DATE 13 Aug 1998

\LONGITUDE 12.0 \LATITUDE -0.1
\RAHMS 18 12 11 \DECDM -18 37
\SIZE 7? \TYPE ?
\FLUX1GHZ 3.5 \ALPHA 0.7
\NAMES

\RADIO Incomplete shell, defined in E only.
\XRAY Detected.
\DATE 10 Oct 2001

\LONGITUDE 13.3 \LATITUDE -1.3
\RAHMS 18 19 20 \DECDM -18 00
\SIZE 70\X/40 \TYPE S?
\FLUX1GHZ ? \ALPHA ?
\NAMES

\RADIO Amorphous emission.
\XRAY Elongated emission.
\OPTICAL Filaments in S.
\DISTANCE Absorption indicates 2--4 kpc.
\DATE 1 Aug 2000

\LONGITUDE 13.5 \LATITUDE +0.2
\RAHMS 18 14 14 \DECDM -17 12
\SIZE 5\X/4 \TYPE S
\FLUX1GHZ 3.5? \ALPHA 1.0?
\NAMES

\NOTES Has been called G13.46$+$0.16.
\RADIO Elongated, incomplete shell.
\DATE 13 Aug 1998

\LONGITUDE 15.1 \LATITUDE -1.6
\RAHMS 18 24 00 \DECDM -16 34
\SIZE 30\X/24 \TYPE S
\FLUX1GHZ 5.5? \ALPHA 0.8?
\NAMES

\RADIO Elongated, incomplete shell.
\DATE 19th May 1992

\LONGITUDE 15.9 \LATITUDE +0.2
\RAHMS 18 18 52 \DECDM -15 02
\SIZE 7\X/5 \TYPE S?
\FLUX1GHZ 5 \ALPHA 0.6?
\NAMES

\RADIO Incomplete shell, with bright concentration to the E.
\DATE 13 Aug 1998

\LONGITUDE 16.2 \LATITUDE -2.7
\RAHMS 18 28 50 \DECDM -16 11
\SIZE 17 \TYPE S
\FLUX1GHZ 2 \ALPHA 0.5
\NAMES

\RADIO Double rim.
\DATE 26 Jul 2000

\LONGITUDE 16.7 \LATITUDE +0.1
\RAHMS 18 20 56 \DECDM -14 20
\SIZE 4 \TYPE C
\FLUX1GHZ 3.0 \ALPHA 0.6
\NAMES

\NOTES Has been called G16.73$+$0.08.
\RADIO Asymmetric shell with flat-spectrum core.
\XRAY Non-thermal core.
\DATE 11 Jan 2004

\LONGITUDE 16.8 \LATITUDE -1.1
\RAHMS 18 25 20 \DECDM -14 46
\SIZE 30\X/24? \TYPE ?
\FLUX1GHZ 2? \ALPHA ?
\NAMES

\NOTES Has been called G16.85$-$1.05.
\RADIO Overlapping thermal and non-thermal emission, parameters uncertain.
\POINT Pulsar within boundary of non-thermal emission.
\DATE 13 Aug 1998

\LONGITUDE 17.4 \LATITUDE -2.3
\RAHMS 18 30 55 \DECDM -14 52
\SIZE 24? \TYPE S
\FLUX1GHZ 4.8? \ALPHA 0.8?
\NAMES

\RADIO Incomplete, poorly defined shell.
\OPTICAL Filaments to SE, and diffuse emission.
\DATE 7 Jan 2004

\LONGITUDE 17.8 \LATITUDE -2.6
\RAHMS 18 32 50 \DECDM -14 39
\SIZE 24 \TYPE S
\FLUX1GHZ 4.0? \ALPHA 0.3?
\NAMES

\RADIO Well defined shell.
\DATE 13 Aug 1998

\LONGITUDE 18.8 \LATITUDE +0.3
\RAHMS 18 23 58 \DECDM -12 23
\SIZE 17\X/11 \TYPE S
\FLUX1GHZ 33 \ALPHA 0.4
\NAMES Kes 67

\NOTES Has been called G18.9$+$0.3.
\RADIO Incomplete shell, in complex region near the H\II/ region W39.
\DISTANCE H\I/ absorption indicates $>9.5$~kpc, and possibly $<19$~kpc.
  plus CO observations.
\DATE 1 Aug 2000

\LONGITUDE 18.9 \LATITUDE -1.1
\RAHMS 18 29 50 \DECDM -12 58
\SIZE 33 \TYPE C?
\FLUX1GHZ 37 \ALPHA varies
\NAMES

\NOTES Has been called G18.95$-$1.1 and G18.94$-$1.04.
\RADIO Non-thermal, diffuse partially limb-brightened, with central ridge.
\XRAY Partial shell.
  ($4'.4\X/4'.1: S=23\PM/6$~Jy).
  ($19''\X/14''$), and Effelsberg 100-m at 1.4~GHz ($9'$) for H\I/.
\DATE 12 Oct 2001

\LONGITUDE 20.0 \LATITUDE -0.2
\RAHMS 18 28 07 \DECDM -11 35
\SIZE 10 \TYPE F
\FLUX1GHZ 10 \ALPHA 0.0
\NAMES

\RADIO Faint, filled-centre, polarized.
\POINT OH source 20.1$-$0.1 is nearby.
\DATE 13 Aug 1998

\LONGITUDE 21.5 \LATITUDE -0.9
\RAHMS 18 33 33 \DECDM -10 35
\SIZE 4 \TYPE C
\FLUX1GHZ 6? \ALPHA 0.0
\NAMES

\NOTES Early observations relate to the central core only.
\RADIO Filled-centre, with high frequency turnover.
\XRAY Central core, with extended, faint halo.
\DISTANCE H\I/ absorption indicates 5.5~kpc or more.
  review of flux densities.
\DATE 7 Jan 2004

\LONGITUDE 21.8 \LATITUDE -0.6
\RAHMS 18 32 45 \DECDM -10 08
\SIZE 20 \TYPE S
\FLUX1GHZ 69 \ALPHA 0.5
\NAMES Kes 69

\RADIO Incomplete shell.
\XRAY Detected.
\DISTANCE H$_2$CO absorption indicates $>6.3$~kpc.
\DATE 27 Jan 2004

\LONGITUDE 22.7 \LATITUDE -0.2
\RAHMS 18 33 15 \DECDM -09 13
\SIZE 26 \TYPE S?
\FLUX1GHZ 33 \ALPHA 0.6
\NAMES

\RADIO Non-thermal ring in complex region, overlapping G23.3$-$0.3.
\DATE 13 Aug 1998

\LONGITUDE 23.3 \LATITUDE -0.3
\RAHMS 18 34 45 \DECDM -08 48
\SIZE 27 \TYPE S
\FLUX1GHZ 70 \ALPHA 0.5
\NAMES W41

\RADIO Incomplete ring, in complex region, overlapping G22.7$-$0.2.
\POINT Pulsar association suggested.
\DATE 13 Aug 1998

\LONGITUDE 23.6 \LATITUDE +0.3
\RAHMS 18 33 03 \DECDM -08 13
\SIZE 10? \TYPE ?
\FLUX1GHZ 8? \ALPHA 0.3
\NAMES

\RADIO Not well resolved, in complex region.
\DATE 13 Aug 1998

\LONGITUDE 24.7 \LATITUDE -0.6
\RAHMS 18 38 43 \DECDM -07 32
\SIZE 15? \TYPE S?
\FLUX1GHZ 8 \ALPHA 0.5
\NAMES

\RADIO Incomplete shell, defined in SW.
\DATE 13 Aug 1998

\LONGITUDE 24.7 \LATITUDE +0.6
\RAHMS 18 34 10 \DECDM -07 05
\SIZE 30\X/15 \TYPE C?
\FLUX1GHZ 20? \ALPHA 0.2?
\NAMES

\RADIO Filled-centre, with faint shell, and a compact H\II/ region to the S.
  NRO 45-m at 10.2~GHz ($2'.7: S=15\PM/3$~Jy).
\DATE 13 Aug 1998

\LONGITUDE 27.4 \LATITUDE +0.0
\RAHMS 18 41 19 \DECDM -04 56
\SIZE 4 \TYPE S
\FLUX1GHZ 6 \ALPHA 0.68
\NAMES 4C$-$04.71

\NOTES Early references refer to G27.3$-$0.1 (Kes 73), a supposed larger remnant.
\RADIO Incomplete shell.
\XRAY Diffuse emission, with central low period pulsar.
\POINT Central AXP.
\DISTANCE H\I/ absorption indicates 6 to 7.5~kpc.
  $S_{\rm 5~GHz}=1.9\PM/0.2$~Jy.
\DATE 17 Oct 2001

\LONGITUDE 27.8 \LATITUDE +0.6
\RAHMS 18 39 50 \DECDM -04 24
\SIZE 50\X/30 \TYPE F
\FLUX1GHZ 30 \ALPHA varies
\NAMES

\RADIO Filled-centre, with spectral turnover.
  NRO 45-m at 10.2~GHz (smoothed to $4'.3: S=8.5\PM/2$~Jy).
\DATE 13 Aug 1998

\LONGITUDE 28.6 \LATITUDE -0.1
\RAHMS 18 43 55 \DECDM -03 53
\SIZE 13\X/9 \TYPE S
\FLUX1GHZ 3? \ALPHA ?
\NAMES

\RADIO Poorly defined regions of non-thermal emission.
\XRAY Diffuse shell, with thermal and non-thermal emission.
\DATE 11 Jan 2004

\LONGITUDE 28.8 \LATITUDE +1.5
\RAHMS 18 39 00 \DECDM -02 55
\SIZE 100? \TYPE S?
\FLUX1GHZ ? \ALPHA 0.4?
\NAMES

\RADIO Part of rim detected.
\XRAY Diffuse, Centrally brightened.
\DATE 17 Oct 2001

\LONGITUDE 29.6 \LATITUDE +0.1
\RAHMS 18 44 52 \DECDM -02 57
\SIZE 5 \TYPE S
\FLUX1GHZ 1.5? \ALPHA 0.5?
\NAMES

\RADIO Diffuse shell.
\POINT AXP associated.
\DATE 10 Oct 2001

\LONGITUDE 29.7 \LATITUDE -0.3
\RAHMS 18 46 25 \DECDM -02 59
\SIZE 3 \TYPE C
\FLUX1GHZ 10 \ALPHA 0.7
\NAMES Kes 75

\RADIO Shell with flatter spectrum emission from centre.
\XRAY Thermal shell and non-thermal core, and central pulsar.
\POINT X-ray pulsar.
\DISTANCE H\I/ absorption indicates $>9$~kpc and possibly at 21~kpc.
  of flux densities.
\DATE 11 Jan 2004

\LONGITUDE 30.7 \LATITUDE -2.0
\RAHMS 18 54 25 \DECDM -02 54
\SIZE 16 \TYPE ?
\FLUX1GHZ 0.5? \ALPHA 0.7?
\NAMES

\RADIO Poorly defined.
\DATE 13 Aug 1998

\LONGITUDE 30.7 \LATITUDE +1.0
\RAHMS 18 44 00 \DECDM -01 32
\SIZE 24\X/18 \TYPE S?
\FLUX1GHZ 6 \ALPHA 0.4
\NAMES

\RADIO Non-thermal, highly polarized part shell?
\POINT Compact source near centre.
\DATE 13 Aug 1998

\LONGITUDE 31.5 \LATITUDE -0.6
\RAHMS 18 51 10 \DECDM -01 31
\SIZE 18? \TYPE S?
\FLUX1GHZ 2? \ALPHA ?
\NAMES

\NOTES Has been called G31.55$-$0.65.
\RADIO Distorted shell? near H\II/ region.
\OPTICAL Diffuse, incomplete shell.
\DATE 17 Oct 2001

\LONGITUDE 31.9 \LATITUDE +0.0
\RAHMS 18 49 25 \DECDM -00 55
\SIZE 7\X/5 \TYPE S
\FLUX1GHZ 24 \ALPHA 0.55
\NAMES 3C391

\RADIO Shell, brightest in NW.
\XRAY Diffuse with central core.
\DISTANCE H\I/ absorption is seen to the tangent point (8.5~kpc).
\DATE 23 Jan 2004

\LONGITUDE 32.0 \LATITUDE -4.9
\RAHMS 19 06 00 \DECDM -03 00
\SIZE 60? \TYPE S?
\FLUX1GHZ 22? \ALPHA 0.5?
\NAMES 3C396.1

\RADIO Possible large shell?
\DATE 25th February 1988

\LONGITUDE 32.1 \LATITUDE -0.9
\RAHMS 18 53 10 \DECDM -01 08
\SIZE 40? \TYPE C?
\FLUX1GHZ ? \ALPHA ?
\NAMES

\RADIO Possible faint shell, not well defined.
\XRAY Diffuse, with clumps.
\DATE 17 Aug 1998

\LONGITUDE 32.8 \LATITUDE -0.1
\RAHMS 18 51 25 \DECDM -00 08
\SIZE 17 \TYPE S?
\FLUX1GHZ 11? \ALPHA 0.2?
\NAMES Kes 78

\NOTES Part has been called G33.1$-$0.1.
\RADIO Elongated shell?
\DATE 1 Aug 2000

\LONGITUDE 33.2 \LATITUDE -0.6
\RAHMS 18 53 50 \DECDM -00 02
\SIZE 18 \TYPE S
\FLUX1GHZ 3.5 \ALPHA varies
\NAMES

\RADIO Incomplete shell.
\DATE 13 Aug 1998

\LONGITUDE 33.6 \LATITUDE +0.1
\RAHMS 18 52 48 \DECDM +00 41
\SIZE 10 \TYPE S
\FLUX1GHZ 22 \ALPHA 0.5
\NAMES Kes 79, 4C00.70, HC13

\NOTES Has been called G33.7$+$0.0.
\RADIO Shell, with bright central region, in complex region.
\XRAY Centrally brightened.
\POINT Central X-ray source.
\DISTANCE H\I/ absorption gives about 10~kpc.
  spectral comparison.
\DATE 23 Jan 2004

\LONGITUDE 34.7 \LATITUDE -0.4
\RAHMS 18 56 00 \DECDM +01 22
\SIZE 35\X/27 \TYPE C
\FLUX1GHZ 230 \ALPHA 0.30
\NAMES W44, 3C392

\NOTES Has been called G34.6$-$0.5.
\RADIO Distorted shell, brighter to the E, with pulsar and associated nebula.
\OPTICAL Diffuse emission.
\XRAY Centrally concentrated, thermal spectrum, plus pulsar wind nebula.
\POINT Pulsar within the boundary of the remnant.
\DISTANCE H\I/ absorption indicates 2.5~kpc.
\DATE 27 Jan 2004

\LONGITUDE 36.6 \LATITUDE -0.7
\RAHMS 19 00 35 \DECDM +02 56
\SIZE 25? \TYPE S?
\FLUX1GHZ ? \ALPHA ?
\NAMES

\RADIO Polarized arc, possibly part of a larger shell?
\DATE 13 Aug 1998

\LONGITUDE 36.6 \LATITUDE +2.6
\RAHMS 18 48 49 \DECDM +04 26
\SIZE 17\X/13? \TYPE S
\FLUX1GHZ 0.7? \ALPHA 0.5?
\NAMES

\RADIO Poorly resolved shell.
\DATE 19th May 1992

\LONGITUDE 39.2 \LATITUDE -0.3
\RAHMS 19 04 08 \DECDM +05 28
\SIZE 8\X/6 \TYPE C
\FLUX1GHZ 18 \ALPHA 0.6
\NAMES 3C396, HC24, NRAO 593

\RADIO Shell, brighter to W, with faint `tail' to E.
\XRAY Diffuse, brighter to W, with central core.
\POINT Central X-ray source.
\DISTANCE H\I/ absorption suggests at least 7.7~kpc.
  polarization, plus Ooty at 327~MHz ($100''\X/31''$), including review of flux densities.
\DATE 11 Jan 2004

\LONGITUDE 39.7 \LATITUDE -2.0
\RAHMS 19 12 20 \DECDM +04 55
\SIZE 120\X/60 \TYPE ?
\FLUX1GHZ 85? \ALPHA 0.7?
\NAMES W50, SS433

\NOTES Eastern part has been called G40.0$-$3.1. Is this a SNR?
\RADIO Elongated shell, containing SS433, adjacent to the H\II/ region S74.
\OPTICAL Faint filaments at the edge of the radio emission.
\XRAY Emission from SS433 and two lobes.
\POINT SS433 is the compact source in the centre of the W50.
\DISTANCE Distance to SS433 is 5~kpc.
  2.7~GHz data.
  plus NRAO 140-ft at 1.4~GHz ($21'$) for H\I/ observations.
\DATE 17 Oct 2001

\LONGITUDE 40.5 \LATITUDE -0.5
\RAHMS 19 07 10 \DECDM +06 31
\SIZE 22 \TYPE S
\FLUX1GHZ 11 \ALPHA 0.5
\NAMES

\RADIO Shell, brightest to the NE.
  plus review of flux densities.
\DATE 17 Oct 2001

\LONGITUDE 41.1 \LATITUDE -0.3
\RAHMS 19 07 34 \DECDM +07 08
\SIZE 4.5\X/2.5 \TYPE S
\FLUX1GHZ 22 \ALPHA 0.48
\NAMES 3C397

\RADIO 3C397 is two sources: the E is the SNR, the W is a H\II/ region.
\XRAY Brighter to the E and W, with possible central component.
\DISTANCE Possible limit of $>7.5$~kpc for non-thermal component from H\I/ absorption.
  and Haystack 36-m at 15.5~GHz ($2'.3: S=8.5\PM/3.0$~Jy).
  polarization and comparison with ROSAT image.
\DATE 27 Jan 2004

\LONGITUDE 42.8 \LATITUDE +0.6
\RAHMS 19 07 20 \DECDM +09 05
\SIZE 24 \TYPE S
\FLUX1GHZ 3? \ALPHA 0.5?
\NAMES

\NOTES Has been called G42.8$+$0.65.
\RADIO Faint shell.
\POINT Near soft gamma repeater, and young pulsar.
\DATE 11 Jan 2004

\LONGITUDE 43.3 \LATITUDE -0.2
\RAHMS 19 11 08 \DECDM +09 06
\SIZE 4\X/3 \TYPE S
\FLUX1GHZ 38 \ALPHA 0.48
\NAMES W49B

\RADIO Shell, brightest to the SE and W, near the H\II/ region W49A.
\XRAY Filled-centre.
\DISTANCE H\I/ absorption indicates 12.5 to 14~kpc.
  ($4''.8\X/5''.2: S=31.8$~Jy) and 4.85~GHz ($4''.0\X/4''.1$), including polarization.
\DATE 23 Jan 2004

\LONGITUDE 43.9 \LATITUDE +1.6
\RAHMS 19 05 50 \DECDM +10 30
\SIZE 60? \TYPE S?
\FLUX1GHZ 8.6? \ALPHA 0.2?
\NAMES

\RADIO Large, poorly defined faint shell.
\POINT Soft gamma repeater nearby.
\DATE 11 Jan 2004

\LONGITUDE 45.7 \LATITUDE -0.4
\RAHMS 19 16 25 \DECDM +11 09
\SIZE 22 \TYPE S
\FLUX1GHZ 4.2? \ALPHA 0.4?
\NAMES

\RADIO Shell, brightest to the SE, poorly defined to NW.
\DATE 13 Aug 1998

\LONGITUDE 46.8 \LATITUDE -0.3
\RAHMS 19 18 10 \DECDM +12 09
\SIZE 17\X/13 \TYPE S
\FLUX1GHZ 14 \ALPHA 0.5
\NAMES (HC30)

\NOTES Has been called G46.6$-$0.2.
\RADIO Shell, two bright arcs to NNW and SSE.
\DISTANCE H\I/ absorption suggests 6.8--8.8~kpc.
\DATE 14 Aug 1998

\LONGITUDE 49.2 \LATITUDE -0.7
\RAHMS 19 23 50 \DECDM +14 06
\SIZE 30 \TYPE S?
\FLUX1GHZ 160? \ALPHA 0.3?
\NAMES (W51)

\RADIO In complex region, parameters uncertain.
\XRAY Elongated east--west.
\OPTICAL Some diffuse emission possibly associated.
\DISTANCE H\I/ absorption suggests 4.1~kpc.
  but probably confused.
\DATE 7 Jan 2004

\LONGITUDE 53.6 \LATITUDE -2.2
\RAHMS 19 38 50 \DECDM +17 14
\SIZE 33\X/28 \TYPE S
\FLUX1GHZ 8 \ALPHA 0.75
\NAMES 3C400.2, NRAO 611

\NOTES Has been called G53.7$-$2.2.
\RADIO Ring of emission, with extension to NW.
\OPTICAL Filaments and diffuse emission.
\XRAY Centrally brightened, offset to NW.
\DISTANCE Mean optical velocity indicates 6.7~kpc, H\I/ absorption indicates 2.3~kpc.
\DATE 17 Oct 2001

\LONGITUDE 54.1 \LATITUDE +0.3
\RAHMS 19 30 31 \DECDM +18 52
\SIZE 1.5 \TYPE F?
\FLUX1GHZ 0.5 \ALPHA 0.1
\NAMES

\RADIO Filled-centre.
\XRAY Centrally concentrated, with extensions and diffuse emission.
\POINT Central pulsar.
  5~GHz ($5'': S=0.33\PM/0.02$~Jy), Ooty at 327~MHz ($S=0.50\PM/0.08$~Jy), plus review of flux densities.
\DATE 7 Jan 2004

\LONGITUDE 54.4 \LATITUDE -0.3
\RAHMS 19 33 20 \DECDM +18 56
\SIZE 40 \TYPE S
\FLUX1GHZ 28 \ALPHA 0.5
\NAMES (HC40)

\NOTES Has been called G54.5$-$0.3.
\RADIO Shell, in complex region.
\DATE 19 Aug 1998

\LONGITUDE 55.0 \LATITUDE +0.3
\RAHMS 19 32 00 \DECDM +19 50
\SIZE 20\X/15? \TYPE S
\FLUX1GHZ 0.5? \ALPHA 0.5?
\NAMES

\NOTES Has been called G55.2$+$0.5.
\RADIO Faint, partial shell.
\DISTANCE Association with H\I/ features implies 14~kpc.
\POINT Old pulsar nearby.
  ($1'.0\X/2'.9: S=0.25\PM/0.12$~Jy), plus H\I/ observations.
\DATE 18 Sep 1998

\LONGITUDE 55.7 \LATITUDE +3.4
\RAHMS 19 21 20 \DECDM +21 44
\SIZE 23 \TYPE S
\FLUX1GHZ 1.4 \ALPHA 0.6
\NAMES

\RADIO Incomplete shell.
\POINT Old pulsar within the boundary of the remnant.
  ($27''\X/72'': S=1.0\PM/0.1$~Jy).
\DATE 05 Apr 1993

\LONGITUDE 57.2 \LATITUDE +0.8
\RAHMS 19 34 59 \DECDM +21 57
\SIZE 12? \TYPE S?
\FLUX1GHZ 1.8? \ALPHA ?
\NAMES (4C21.53)

\RADIO Extended non-thermal arc.
\POINT Near the millisecond pulsar, but not thought to be related.
\DATE 1st March 1988

\LONGITUDE 59.5 \LATITUDE +0.1
\RAHMS 19 42 33 \DECDM +23 35
\SIZE 5 \TYPE S
\FLUX1GHZ  3?  \ALPHA  ?
\NAMES

\NOTES Has been called G59.6$+$0.1.
\RADIO Incomplete shell.
\DATE 14 Aug 1998

\LONGITUDE 59.8 \LATITUDE +1.2
\RAHMS 19 38 55 \DECDM +24 19
\SIZE 20\X/16? \TYPE ?
\FLUX1GHZ 1.6 \ALPHA 0.5
\NAMES

\NOTES Has been called G59.7$+$1.2.
\RADIO Poorly defined source.
\DATE 28 Jun 1995

\LONGITUDE 63.7 \LATITUDE +1.1
\RAHMS 19 47 52 \DECDM +27 45
\SIZE 8 \TYPE F
\FLUX1GHZ 1.8 \ALPHA 0.3
\NAMES

\RADIO Centrally brightened, with core.
\DATE 13 Aug 1998

\LONGITUDE 65.1 \LATITUDE +0.6
\RAHMS 19 54 40 \DECDM +28 35
\SIZE 90\X/50 \TYPE S
\FLUX1GHZ 6 \ALPHA 0.6
\NAMES

\RADIO Large, faint shell.
\POINT Pulsar nearby.
\DATE 13 Aug 1998

\LONGITUDE 65.3 \LATITUDE +5.7
\RAHMS 19 33 00 \DECDM +31 10
\SIZE 310\X/240 \TYPE S?
\FLUX1GHZ 52? \ALPHA 0.6?
\NAMES

\NOTES Has been called G65.2$+$5.7.
\RADIO Large, faint ring? near S91 and S94.
\OPTICAL Filamentary ring.
\XRAY Detected.
\DISTANCE Mean optical velocity suggests 0.8~kpc.
\DATE 7 Jan 2004

\LONGITUDE 65.7 \LATITUDE +1.2
\RAHMS 19 52 10 \DECDM +29 26
\SIZE 18 \TYPE ?
\FLUX1GHZ 5.1 \ALPHA 0.6
\NAMES DA 495

\NOTES Has mistakenly been called G55.7$+$1.2.
\RADIO Filled-centre or thick shell?
  plus review of flux densities.
  1.4~GHz ($36''$ and $12''$), including IRAS imaging.
\DATE 12 Oct 2001

\LONGITUDE 67.7 \LATITUDE +1.8
\RAHMS 19 54 32 \DECDM +31 29
\SIZE  9 \TYPE  S
\FLUX1GHZ  1.4 \ALPHA  0.3
\NAMES

\RADIO Double arc shell.
\OPTICAL Filaments in N.
\DATE 17 Oct 2001

\LONGITUDE 68.6 \LATITUDE -1.2
\RAHMS 20 08 40 \DECDM +30 37
\SIZE 28\X/25? \TYPE ?
\FLUX1GHZ 0.7? \ALPHA 0.0?
\NAMES

\RADIO Faint, poorly defined source.
\DATE 13 Aug 1998

\LONGITUDE 69.0 \LATITUDE +2.7
\RAHMS 19 53 20 \DECDM +32 55
\SIZE 80? \TYPE ?
\FLUX1GHZ 120? \ALPHA varies
\NAMES CTB 80

\NOTES An association with a SN in \AD/1408 has been suggested. Has been called G68.8$+$2.8. Is it a SNR?
\RADIO Compact core, flat spectrum plateau, and steeper spectrum extensions, with spectral break?
\OPTICAL Expanding nebulosity near centre, with filaments to the SW and far NE.
\XRAY Diffuse emission with compact source.
\POINT Pulsar at western edge of core.
\DATE 11 Jan 2004

\LONGITUDE 69.7 \LATITUDE +1.0
\RAHMS 20 02 40 \DECDM +32 43
\SIZE 16 \TYPE S
\FLUX1GHZ 1.6 \ALPHA 0.8
\NAMES

\RADIO Poorly resolved source.
\DATE 13 Aug 1998

\LONGITUDE 73.9 \LATITUDE +0.9
\RAHMS 20 14 15 \DECDM +36 12
\SIZE 22? \TYPE S?
\FLUX1GHZ 9? \ALPHA 0.3?
\NAMES

\RADIO Diffuse, centrally brightened to SW.
\OPTICAL Faint shell.
\DATE 7 Jan 2004

\LONGITUDE 74.0 \LATITUDE -8.5
\RAHMS 20 51 00 \DECDM +30 40
\SIZE 230\X/160 \TYPE S
\FLUX1GHZ 210 \ALPHA varies
\NAMES Cygnus Loop

\NOTES Has been suggested that this is two overlapping remnants.
\RADIO Shell, brightest to the NE, with fainter breakout region to S, with spectral variations.
\OPTICAL Large filamentary loop, brightest to the NE, not well defined to the S or W.
\XRAY Shell in soft X-rays.
\POINT Several compact radio sources within the boundary of the remnant, including CL4, plus X-ray sources in S.
\DISTANCE Optical proper motion and shock velocity gives 0.44~kpc.
  dominated filaments in NE.
  index studies in comparison with other radio observations.
\DATE 11 Jan 2004

\LONGITUDE 74.9 \LATITUDE +1.2
\RAHMS 20 16 02 \DECDM +37 12
\SIZE 8\X/6 \TYPE F
\FLUX1GHZ 9 \ALPHA varies
\NAMES CTB 87

\RADIO Filled-centre, with high polarization and high frequency turnover.
\XRAY Centrally brightened.
\DISTANCE H\I/ absorption indicates 12~kpc, optical extinction gives $6.1\PM/0.9$~kpc.
\POINT Extragalactic compact source is nearby.
\DATE 11 Jan 2004

\LONGITUDE 76.9 \LATITUDE +1.0
\RAHMS 20 22 20 \DECDM +38 43
\SIZE 12\X/9 \TYPE ?
\FLUX1GHZ 2 \ALPHA 0.6
\NAMES

\RADIO Diffuse, non-thermal, with low frequency turnover.
  8.55~GHz ($11''\X/12''$), including polarization and review of flux densities.
\DATE 13 Aug 1998

\LONGITUDE 78.2 \LATITUDE +2.1
\RAHMS 20 20 50 \DECDM +40 26
\SIZE 60 \TYPE S
\FLUX1GHZ 340 \ALPHA 0.5
\NAMES DR4, $\gamma$ Cygni SNR

\NOTES Has been called G78.1$+$1.8.
\RADIO In complex region (early catalogues refer to other proposed remnants in this region).
\OPTICAL Faint filaments, spectra indicate a SNR superposed on a H\II/ region.
\XRAY Weak emission from the SE of the remnant.
\POINT $\gamma$-ray and X-ray point source in remnant.
\DATE 7 Jan 2004

\LONGITUDE 82.2 \LATITUDE +5.3
\RAHMS 20 19 00 \DECDM +45 30
\SIZE 95\X/65 \TYPE S
\FLUX1GHZ 120? \ALPHA 0.5?
\NAMES W63

\NOTES Has been called G82.5$+$5.3.
\RADIO Shell in the Cygnus X complex.
\OPTICAL In complex region, but spectra indicate SNR filaments.
\XRAY Detected.
\DATE 11 Jan 2004

\LONGITUDE 84.2 \LATITUDE -0.8
\RAHMS 20 53 20 \DECDM +43 27
\SIZE 20\X/16 \TYPE S
\FLUX1GHZ 11 \ALPHA 0.5
\NAMES

\RADIO Elongated shell, with a filament aligned with the major axis.
\DATE 11 Jan 2004

\LONGITUDE 84.9 \LATITUDE +0.5
\RAHMS 20 50 30 \DECDM +44 53
\SIZE  6 \TYPE  S
\FLUX1GHZ  0.8 \ALPHA  0.4
\NAMES

\RADIO Incomplete shell.
\DATE 13 Aug 1998

\LONGITUDE 85.4 \LATITUDE +0.7
\RAHMS 20 50 40 \DECDM +45 22
\SIZE 24 \TYPE S
\FLUX1GHZ ? \ALPHA 0.5?
\NAMES

\RADIO Faint, incomplete shell, within larger thermal shell.
\XRAY Detected.
     ($1'.1\X/0'.8$), plus H\I/, X-ray and optical data.
\DATE 12 Oct 2001

\LONGITUDE 85.9 \LATITUDE -0.6
\RAHMS 20 58 40 \DECDM +44 53
\SIZE 24 \TYPE S
\FLUX1GHZ ? \ALPHA 0.5?
\NAMES

\RADIO Faint, incomplete shell.
\XRAY Detected.
     ($1'.1\X/0'.8$), plus H\I/, X-ray and optical data.
\DATE 12 Oct 2001

\LONGITUDE 89.0 \LATITUDE +4.7
\RAHMS 20 45 00 \DECDM +50 35
\SIZE 120\X/90 \TYPE S
\FLUX1GHZ 220 \ALPHA 0.40
\NAMES HB21

\RADIO Distorted shell (4C50.52, an extragalactic double, is within the boundary of the remnant).
\OPTICAL Filaments possibly associated.
\XRAY Centrally brightened.
\DISTANCE Various associations suggest 0.8~kpc.
  plus CO observations of adjacent molecular cloud.
\DATE 11 Jan 2004

\LONGITUDE 93.3 \LATITUDE +6.9
\RAHMS 20 52 25 \DECDM +55 21
\SIZE 27\X/20 \TYPE S
\FLUX1GHZ 9 \ALPHA 0.54
\NAMES DA 530, 4C(T)55.38.1

\NOTES Has been called G93.2$+$6.7.
\RADIO Shell, with two bright limbs, highly polarized.
\DISTANCE H\I/ observations suggest 2.2~kpc.
\DATE 11 Jan 2004

\LONGITUDE 93.7 \LATITUDE -0.2
\RAHMS 21 29 20 \DECDM +50 50
\SIZE 80 \TYPE S
\FLUX1GHZ 65 \ALPHA 0.4
\NAMES CTB 104A, DA 551

\NOTES Has been called G93.6$-$0.2 and G93.7$-$0.3.
\RADIO Distorted, faint shell.
\DISTANCE Assocation with H\I/ features suggests 1.5~kpc.
  polarization, plus review of flux densities.
\DATE 11 Jan 2004

\LONGITUDE 94.0 \LATITUDE +1.0
\RAHMS 21 24 50 \DECDM +51 53
\SIZE 30\X/25 \TYPE S
\FLUX1GHZ 15 \ALPHA 0.44
\NAMES 3C434.1

\RADIO Incomplete shell, in complex region.
\DATE 11 Jan 2004

\LONGITUDE 106.3 \LATITUDE +2.7
\RAHMS 22 27 30 \DECDM +60 50
\SIZE 60\X/24 \TYPE ?
\FLUX1GHZ 6 \ALPHA 0.6
\NAMES

\NOTES Incorporates the proposed smaller remnant G106.6$+$2.9.
\RADIO Faint extended source, which brighter `head' to NE.
\POINT Pulsar in `head'.
\DATE 26 Nov 2001

\LONGITUDE 109.1 \LATITUDE -1.0
\RAHMS 23 01 35 \DECDM +58 53
\SIZE 28 \TYPE S
\FLUX1GHZ 20 \ALPHA 0.50
\NAMES CTB 109

\RADIO Semicircular shell, with the Molecular cloud S152 is to the immediate W.
\XRAY Incomplete shell, with pulsar at W edge.
\POINT Long period X-ray pulsar.
\DISTANCE Assocation with H\II/ regions implies 3~kpc.
\DATE 7 Jan 2004

\LONGITUDE 111.7 \LATITUDE -2.1
\RAHMS 23 23 26 \DECDM +58 48
\SIZE 5 \TYPE S
\FLUX1GHZ 2720 \ALPHA 0.77
\NAMES Cassiopeia A, 3C461

\NOTES Presumably the remnant of a late 17th century SN.
\RADIO Bright shell with compact knots and extended plateau of emission.
\OPTICAL Fast knots and quasi-stationary flocculli, with many filaments at large radii, and NE `jet'.
\XRAY Incomplete shell, with hard spectral component, and compact central source.
\DISTANCE Optical expansion, plus proper motions indicate 3.4~kpc.
  images for spectral index studies.
  plus CO observations.
\DATE 23 Jan 2004

\LONGITUDE 114.3 \LATITUDE +0.3
\RAHMS 23 37 00 \DECDM +61 55
\SIZE 90\X/55 \TYPE S
\FLUX1GHZ 6? \ALPHA 0.3?
\NAMES

\RADIO Shell, with H\II/ region S165 within the boundary of the remnant.
\OPTICAL Faint emission in centre and to S.
\DISTANCE Possible association with H\I/ features suggests 3.0--3.8~kpc.
\POINT Pulsar near centre of remnant.
  data, plus H\I/ from Maryland-Green Bank survey.
\DATE 7 Jan 2004

\LONGITUDE 116.5 \LATITUDE +1.1
\RAHMS 23 53 40 \DECDM +63 15
\SIZE 80\X/60 \TYPE S
\FLUX1GHZ 11? \ALPHA 0.8?
\NAMES

\RADIO Distinct shell, with high polarization.
\DISTANCE Possible association with H\I/ features suggests 3.6--5.2~kpc.
  1.4~GHz survey data, plus H\I/ from Maryland-Green Bank survey.
\DATE 13 Aug 1998

\LONGITUDE 116.9 \LATITUDE +0.2
\RAHMS 23 59 10 \DECDM +62 26
\SIZE 34 \TYPE S
\FLUX1GHZ 9? \ALPHA 0.5?
\NAMES CTB 1

\NOTES Has been called G117.3$+$0.1 and G116.9$+$0.1.
\RADIO Incomplete shell.
\OPTICAL Filaments on sky survey.
\XRAY Centrally brightened, with NE `breakout'.
\POINT Pulsar to NE.
\DISTANCE Possible association with H\I/ features suggests 2.8--4.0~kpc, mean optical velocity suggests 2.7~kpc.
  plus review of flux densities.
  1.4~GHz survey data, plus H\I/ from Maryland-Green Bank survey.
\DATE 13 Aug 1998

\LONGITUDE 119.5 \LATITUDE +10.2
\RAHMS 00 06 40 \DECDM +72 45
\SIZE 90? \TYPE S
\FLUX1GHZ 36 \ALPHA 0.6
\NAMES CTA 1

\NOTES Has been called G119.5$+$10.3.
\RADIO Incomplete shell, with `breakout' to NW.
\OPTICAL Faint diffuse nebulosities.
\XRAY Centrally brightened.
\POINT Compact, central X-/$\gamma$-ray source.
\DATE 8 Nov 2001

\LONGITUDE 120.1 \LATITUDE +1.4
\RAHMS 00 25 18 \DECDM +64 09
\SIZE 8 \TYPE S
\FLUX1GHZ 56 \ALPHA 0.61
\NAMES Tycho, 3C10, SN1572

\NOTES This is the remnant of the Tycho's SN of \AD/1572.
\RADIO Shell, brightest to the NE.
\OPTICAL Faint filaments/knots to the NNW, NE and E.
\XRAY Shell, brighter to the NE.
\POINT Faint radio source near centre of the remnant, thought to be extragalactic.
\DISTANCE H\I/ absorption gives 2--5~kpc, optical proper motion and shock velocity gives 2.4~kpc.
  1980, for expansion.
  neutral hydrogen studies.
\DATE 7 Jan 2004

\LONGITUDE 126.2 \LATITUDE +1.6
\RAHMS 01 22 00 \DECDM +64 15
\SIZE 70 \TYPE S?
\FLUX1GHZ 7 \ALPHA varies
\NAMES

\RADIO Poorly defined shell.
\OPTICAL Filaments detected.
  ($1'.1\X/1'.0$), plus review of flux densities.
\DATE 7 Jan 2004

\LONGITUDE 127.1 \LATITUDE +0.5
\RAHMS 01 28 20 \DECDM +63 10
\SIZE 45 \TYPE S
\FLUX1GHZ 13 \ALPHA 0.6
\NAMES R5

\NOTES Has been called G127.3$+$0.7.
\RADIO Distinct shell, with bright central source.
\POINT Flat radio spectrum (extragalactic) source at centre of remnant.
\OPTICAL Detected.
\DISTANCE 1.2--1.3~kpc if associated with NGC559.
\DATE 7 Jan 2004

\LONGITUDE 130.7 \LATITUDE +3.1
\RAHMS 02 05 41 \DECDM +64 49
\SIZE 9\X/5 \TYPE F
\FLUX1GHZ 33 \ALPHA 0.10
\NAMES 3C58, SN1181

\NOTES This is the remnant of the SN of \AD/1181.
\RADIO Filled-centre, highly polarized, with high frequency turnover.
\OPTICAL Faint filaments.
\XRAY Centrally brightened with power-law spectrum.
\POINT Central pulsar.
\DISTANCE H\I/ absorption indicates 3.2~kpc.
  for limit on shell.
studies, and comparison with earlier observations for expansion studies.
\DATE 7 Jan 2004

\LONGITUDE 132.7 \LATITUDE +1.3
\RAHMS 02 17 40 \DECDM +62 45
\SIZE 80 \TYPE S
\FLUX1GHZ 45 \ALPHA 0.6
\NAMES HB3

\NOTES Has been called G132.4$+$2.2.
\RADIO Faint shell, adjacent to W3/4/5 complex.
\OPTICAL Complete, filamentary shell, shock excited spectra.
\XRAY Partial shell.
\POINT Pulsar nearby.
\DISTANCE Interaction with surroundings suggests 2.2~kpc.
\DATE 7 Jan 2004

\LONGITUDE 156.2 \LATITUDE +5.7
\RAHMS 04 58 40 \DECDM +51 50
\SIZE 110 \TYPE S
\FLUX1GHZ 5 \ALPHA 0.5
\NAMES

\RADIO Faint shell.
\XRAY Faint shell.
  plus H\I/ and IRAS.
\DATE 13 Aug 2000

\LONGITUDE 160.9 \LATITUDE +2.6
\RAHMS 05 01 00 \DECDM +46 40
\SIZE 140\X/120 \TYPE S
\FLUX1GHZ 110 \ALPHA 0.6
\NAMES HB9

\NOTES Has been called G160.5$+$2.8 and G160.4$+$2.8.
\RADIO Large, filamentary shell.
\OPTICAL Incomplete shell.
\XRAY Centrally brightened.
\POINT Pulsar within boundary of the remnant, plus several nearby compact radio sources.
\DISTANCE Various observations suggests less than 4~kpc.
  plus review of flux densities.
  discussion of distance.
  100-m at 4.7~GHz ($2'.5$) for spectral index studies.
\DATE 7 Jan 2004

\LONGITUDE 166.0 \LATITUDE +4.3
\RAHMS 05 26 30 \DECDM +42 56
\SIZE 55\X/35 \TYPE S
\FLUX1GHZ 7? \ALPHA 0.4?
\NAMES VRO 42.05.01

\RADIO Two arcs of strikingly different radii.
\OPTICAL Nearly complete ring.
\XRAY Predominantly in SW.
\DISTANCE H\I/ indicates 4.5~kpc.
\DATE 13 Aug 1998

\LONGITUDE 166.2 \LATITUDE +2.5
\RAHMS 05 19 00 \DECDM +41 55
\SIZE 90\X/70 \TYPE S
\FLUX1GHZ 11 \ALPHA 0.5
\NAMES OA 184

\RADIO Large, faint shell.
\OPTICAL Nearly complete ring.
\DISTANCE H\I/ indicates $8\PM/2$~kpc.
\DATE 13 Aug 1998

\LONGITUDE 179.0 \LATITUDE +2.6
\RAHMS 05 53 40 \DECDM +31 05
\SIZE 70 \TYPE S?
\FLUX1GHZ 7 \ALPHA 0.4
\NAMES

\RADIO Thick shell, with background extragalactic sources near centre.
\DATE 13 Aug 1998

\LONGITUDE 180.0 \LATITUDE -1.7
\RAHMS 05 39 00 \DECDM +27 50
\SIZE 180 \TYPE S
\FLUX1GHZ 65 \ALPHA varies
\NAMES S147

\RADIO Large faint shell, with spectral break.
\OPTICAL Wispy ring.
\XRAY Possible detection.
\POINT Pulsar within boundary, with faint wind nebula.
\DATE 11 Jan 2004

\LONGITUDE 182.4 \LATITUDE +4.3
\RAHMS 06 08 10 \DECDM +29 00
\SIZE 50 \TYPE S
\FLUX1GHZ 1.2 \ALPHA 0.4
\NAMES

\RADIO Incomplete shell.
\DATE 14 Aug 1998

\LONGITUDE 184.6 \LATITUDE -5.8
\RAHMS 05 34 31 \DECDM +22 01
\SIZE 7\X/5 \TYPE F
\FLUX1GHZ 1040 \ALPHA 0.30
\NAMES Crab Nebula, 3C144, SN1054

\NOTES This is the remnant of the SN of \AD/1054.
\RADIO Filled-centre, central pulsar, with faint `jet' (or tube) extending from the N edge.
\OPTICAL Strongly polarized filaments, diffuse synchrotron emission, with `jet' faintly visible.
\XRAY Central `torus' around the pulsar.
\POINT Pulsar powering the remnant.
\DISTANCE Proper motions and radial velocities give 2~kpc.
\DATE 23 Jan 2004

\LONGITUDE 189.1 \LATITUDE +3.0
\RAHMS 06 17 00 \DECDM +22 34
\SIZE 45 \TYPE C
\FLUX1GHZ 160 \ALPHA 0.36
\NAMES IC443, 3C157

\RADIO Limb-brightened to NE, with faint extension to the E.
\OPTICAL Brightest to the NE, with faint filaments outside the NE boundary.
\XRAY Shell, brightest to the NE, with nebula and compact source.
\POINT Compact X-ray source in S.
\DISTANCE Mean optical velocity suggests 0.7--1.5~kpc, association with S249 gives 1.5--2~kpc.
  ($12''\X/31''$).
  plus review of flux densities.
  ($3''.8\X/3''.3$ and $40''$).
\DATE 11 Jan 2004

\LONGITUDE 192.8 \LATITUDE -1.1
\RAHMS 06 09 20 \DECDM +17 20
\SIZE 78 \TYPE S
\FLUX1GHZ 20? \ALPHA 0.6?
\NAMES PKS 0607$+$17

\NOTES Has been called G193.3$-$1.5. Has been regarded as part of the Origem Loop, a supposed larger remnant.
\RADIO In complex region.
\OPTICAL Encompasses S261 and S254--258.
\DATE 31 Dec 2001

\LONGITUDE 205.5 \LATITUDE +0.5
\RAHMS 06 39 00 \DECDM +06 30
\SIZE 220 \TYPE S
\FLUX1GHZ 160 \ALPHA 0.5
\NAMES Monoceros Nebula

\RADIO In complex region, parts may be H\II/ regions.
\OPTICAL Large ring, near Rosette nebula.
\XRAY Possibly detected.
\DISTANCE Mean optical velocity suggests 0.8~kpc, low frequency radio absorption suggests 1.6~kpc.
\DATE 13 Aug 1998

\LONGITUDE 206.9 \LATITUDE +2.3
\RAHMS 06 48 40 \DECDM +06 26
\SIZE 60\X/40 \TYPE S?
\FLUX1GHZ 6 \ALPHA 0.5
\NAMES PKS 0646$+$06

\RADIO Diffuse source near the Monoceros Nebula.
\OPTICAL Filaments detected.
\XRAY Possibly detected.
\DATE 13 Aug 1998

\LONGITUDE 260.4 \LATITUDE -3.4
\RAHMS 08 22 10 \DECDM -43 00
\SIZE 60\X/50 \TYPE S
\FLUX1GHZ 130 \ALPHA 0.5
\NAMES Puppis A, MSH 08$-$4{\sl 4}

\NOTES This remnant overlaps the Vela SNR (G263.9$-$3.3).
\RADIO Angular shell, brightest to the E, poorly defined to the W.
\OPTICAL Nebulosity and wisps.
\XRAY Brightest to the E.
\POINT Central source, a possible pulsar in X-rays.
\DISTANCE Association with H\I/ gives $2.2\PM/0.3$~kpc; OH absorption/emission in vicinty implies 0.5 to 1.9~kpc.
\DATE 11 Jan 2004

\LONGITUDE 261.9 \LATITUDE +5.5
\RAHMS 09 04 20 \DECDM -38 42
\SIZE 40\X/30 \TYPE S
\FLUX1GHZ 10? \ALPHA 0.4?
\NAMES

\RADIO Faint shell with little limb brightening.
\DATE 13 Aug 1998

\LONGITUDE 263.9 \LATITUDE -3.3
\RAHMS 08 34 00 \DECDM -45 50
\SIZE 255 \TYPE C
\FLUX1GHZ 1750 \ALPHA varies
\NAMES Vela (XYZ)

\NOTES This refers to the whole Vela XYZ complex, of which X has at times been classified as a separate (filled-centre) remnant. This remnant is overlapped by G260.4$-$3.4 and G266.2$-$1.2.
\RADIO Large shell, with flatter spectrum component (Vela X), and pulsar nebula.
\OPTICAL Filaments.
\XRAY Patchy shell, with extensions, central nebula and pulsar.
\POINT Pulsar within Vela X, with one-sided `jet'.
\DISTANCE Vela pulsar parallax gives 0.3~kpc, optical spectra and H\I/ studies suggest 0.25~kpc.
\DATE 11 Jan 2004

\LONGITUDE 266.2 \LATITUDE -1.2
\RAHMS 08 52 00 \DECDM -46 20
\SIZE 120 \TYPE S
\FLUX1GHZ 50? \ALPHA 0.3?
\NAMES

\NOTES This remnant overlaps the Vela SNR (G263.9$-$3.3).
\RADIO Incomplete shell, confused by the Vela SNR.
\OPTICAL Nebulosity offset to NE.
\XRAY Non-thermal shell, confused by the Vela SNR, with central source.
\POINT Central X-ray source, with optical nebula.
\DISTANCE X-ray data suggest an upper limit of 1~kpc.
\DATE 11 Jan 2004

\LONGITUDE 272.2 \LATITUDE -3.2
\RAHMS 09 06 50 \DECDM -52 07
\SIZE 15? \TYPE S?
\FLUX1GHZ 0.4 \ALPHA 0.6
\NAMES

\RADIO Diffuse shell.
\XRAY Centrally brightened.
\OPTICAL Detected.
\DATE 7 Jan 2004

\LONGITUDE 279.0 \LATITUDE +1.1
\RAHMS 09 57 40 \DECDM -53 15
\SIZE 95 \TYPE S
\FLUX1GHZ 30? \ALPHA 0.6?
\NAMES

\RADIO Faint, incomplete shell.
\POINT Pulsar nearby.
\DATE 13 Aug 1998

\LONGITUDE 284.3 \LATITUDE -1.8
\RAHMS 10 18 15 \DECDM -59 00
\SIZE 24? \TYPE S
\FLUX1GHZ 11? \ALPHA 0.3?
\NAMES MSH 10$-$5{\sl 3}

\NOTES Has been called G284.2$-$1.8.
\RADIO Incomplete, poorly defined shell.
\POINT Pulsar nearby.
 including polarization, plus earlier flux densities.
\DATE 10 Oct 2001

\LONGITUDE 286.5 \LATITUDE -1.2
\RAHMS 10 35 40 \DECDM -59 42
\SIZE 26\X/6 \TYPE S?
\FLUX1GHZ 1.4? \ALPHA ?
\NAMES

\RADIO Double, elongated arc.
\DATE 13 Aug 1998

\LONGITUDE 289.7 \LATITUDE -0.3
\RAHMS 11 01 15 \DECDM -60 18
\SIZE 18\X/14 \TYPE S
\FLUX1GHZ 6.2 \ALPHA 0.2?
\NAMES

\RADIO Incomplete shell.
\POINT Compact radio source near centre.
\DATE 21 Aug 1996

\LONGITUDE 290.1 \LATITUDE -0.8
\RAHMS 11 03 05 \DECDM -60 56
\SIZE 19\X/14 \TYPE S
\FLUX1GHZ 42 \ALPHA 0.4
\NAMES MSH 11$-$6{\sl 1}A

\RADIO Elongated, clumpy shell.
\OPTICAL Filaments detected.
\XRAY Centrally brightened.
\POINT Pulsar nearby.
\DATE 11 Jan 2004

\LONGITUDE 291.0 \LATITUDE -0.1
\RAHMS 11 11 54 \DECDM -60 38
\SIZE 15\X/13 \TYPE C
\FLUX1GHZ 16 \ALPHA 0.29
\NAMES (MSH 11$-$6{\sl 2})

\RADIO Centrally brightened core, with surrounding arcs.
\XRAY Centrally brightened.
\POINT Central compact X-ray source.
\DATE 13 Aug 1998

\LONGITUDE 292.0 \LATITUDE +1.8
\RAHMS 11 24 36 \DECDM -59 16
\SIZE 12\X/8 \TYPE C
\FLUX1GHZ 15 \ALPHA 0.4
\NAMES MSH 11$-$5{\sl 4}

\RADIO Centrally brightened source surrounded by a plateau of faint emission.
\OPTICAL Oxygen rich.
\XRAY Ring of emission, with diffuse central nebula and pulsar.
\POINT Central pulsar.
\DISTANCE H\I/ absorption implies $6.2\PM/0.9$~kpc.
\DATE 23 Jan 2004

\LONGITUDE 292.2 \LATITUDE -0.5
\RAHMS 11 19 20 \DECDM -61 28
\SIZE 20\X/15 \TYPE S
\FLUX1GHZ 7? \ALPHA 0.6?
\NAMES

\RADIO Shell.
\XRAY Detected.
\POINT Central, young pulsar.
\DATE 23 Jan 2004

\LONGITUDE 293.8 \LATITUDE +0.6
\RAHMS 11 35 00 \DECDM -60 54
\SIZE 20 \TYPE C
\FLUX1GHZ 5? \ALPHA 0.6?
\NAMES

\RADIO Central source, with faint extended plateau.
\DATE 21 Aug 1996

\LONGITUDE 294.1 \LATITUDE -0.0
\RAHMS 11 36 10 \DECDM -61 38
\SIZE 40 \TYPE S
\FLUX1GHZ $>$2? \ALPHA ?
\NAMES

\RADIO Faint shell.
\DATE 21 Aug 1996

\LONGITUDE 296.1 \LATITUDE -0.5
\RAHMS 11 51 10 \DECDM -62 34
\SIZE 37\X/25 \TYPE S
\FLUX1GHZ 8? \ALPHA 0.6?
\NAMES

\NOTES Incorporates the previously catalogued remnant G296.1$-$0.7. Has been called G296.05$-$0.50.
\RADIO Irregular shell, with nearby H\II/ regions.
\OPTICAL Detected.
\XRAY Detected.
  poor 5-GHz map (of G296.1$-$0.7).
\DATE 19 Aug 1998

\LONGITUDE 296.5 \LATITUDE +10.0
\RAHMS 12 09 40 \DECDM -52 25
\SIZE 90\X/65 \TYPE S
\FLUX1GHZ 48 \ALPHA 0.5
\NAMES PKS 1209$-$51/52

\NOTES Has been called G296.5$+$9.7.
\RADIO Shell with two bright limbs.
\OPTICAL Detected.
\XRAY Incomplete shell, with central pulsar.
\POINT Central pulsar.
\DATE 10 Oct 2001

\LONGITUDE 296.8 \LATITUDE -0.3
\RAHMS 11 58 30 \DECDM -62 35
\SIZE 20\X/14 \TYPE S
\FLUX1GHZ 9 \ALPHA 0.6
\NAMES 1156$-$62

\RADIO Shell, brighter to the NW.
\DISTANCE H\I/ absorption gives $9.6\PM/0.6$~kpc.
\XRAY Detected.
  observations, plus review of flux densities.
\DATE 19 Aug 1998

\LONGITUDE 298.5 \LATITUDE -0.3
\RAHMS 12 12 40 \DECDM -62 52
\SIZE 5? \TYPE ?
\FLUX1GHZ 5? \ALPHA 0.4?
\NAMES

\RADIO Not well resolved, may be part of a larger ring?
\DATE 21 Aug 1996

\LONGITUDE 298.6 \LATITUDE -0.0
\RAHMS 12 13 41 \DECDM -62 37
\SIZE 12\X/9 \TYPE S
\FLUX1GHZ 5? \ALPHA 0.3
\NAMES

\NOTES Has been called G298.6$-$0.1.
\RADIO Incomplete shell, in complex region.
\DATE 21 Aug 1996

\LONGITUDE 299.2 \LATITUDE -2.9
\RAHMS 12 15 13 \DECDM -65 30
\SIZE 18\X/11 \TYPE S
\FLUX1GHZ 0.5? \ALPHA ?
\NAMES

\RADIO Faint source.
\XRAY Centrally brightened.
\OPTICAL Filaments in W.
\DATE 23 Aug 2000

\LONGITUDE 299.6 \LATITUDE -0.5
\RAHMS 12 21 45 \DECDM -63 09
\SIZE 13 \TYPE S
\FLUX1GHZ 1.0? \ALPHA ?
\NAMES

\RADIO Faint shell, brightest to E.
\DATE 21 Aug 1996

\LONGITUDE 301.4 \LATITUDE -1.0
\RAHMS 12 37 55 \DECDM -63 49
\SIZE 37\X/23 \TYPE S
\FLUX1GHZ 2.1? \ALPHA ?
\NAMES

\RADIO Faint, incomplete shell, with possible extensionm to southwest.
\DATE 21 Aug 1996

\LONGITUDE 302.3 \LATITUDE +0.7
\RAHMS 12 45 55 \DECDM -62 08
\SIZE 17 \TYPE S
\FLUX1GHZ 5? \ALPHA 0.4?
\NAMES

\RADIO Distorted shell, in complex region, with possibly associated filament.
\DATE 21 Aug 1996

\LONGITUDE 304.6 \LATITUDE +0.1
\RAHMS 13 05 59 \DECDM -62 42
\SIZE 8 \TYPE S
\FLUX1GHZ 14 \ALPHA 0.5
\NAMES Kes 17

\RADIO Incomplete shell.
\DISTANCE Possible limit of $>9.7$~kpc from H\I/ absorption.
\DATE 21 Aug 1996

\LONGITUDE 308.1 \LATITUDE -0.7
\RAHMS 13 37 37 \DECDM -63 04
\SIZE 13 \TYPE S
\FLUX1GHZ 1.2? \ALPHA ?
\NAMES

\RADIO Faint shell.
\DATE 21 Aug 1996

\LONGITUDE 308.8 \LATITUDE -0.1
\RAHMS 13 42 30 \DECDM -62 23
\SIZE 30\X/20? \TYPE C?
\FLUX1GHZ 15? \ALPHA 0.4?
\NAMES

\NOTES Incorporates previous catalogued remnant G308.7$+$0.0.
\RADIO Bright ridge in north, and arc to south.
\POINT Pulsar near centre of remnant.
\DATE 13 Aug 1998

\LONGITUDE 309.2 \LATITUDE -0.6
\RAHMS 13 46 31 \DECDM -62 54
\SIZE 15\X/12 \TYPE S
\FLUX1GHZ 7? \ALPHA 0.4?
\NAMES

\NOTES Has been called G309.2$-$0.7.
\RADIO Distorted shell.
\XRAY Extenmed emission, with central source.
\DATE 10 Oct 2001

\LONGITUDE 309.8 \LATITUDE +0.0
\RAHMS 13 50 30 \DECDM -62 05
\SIZE 25\X/19 \TYPE S
\FLUX1GHZ 17 \ALPHA 0.5
\NAMES

\RADIO Distorted shell.
\POINT Steep radio spectrum source near the centre of the remnant.
\DATE 13 Aug 1998

\LONGITUDE 310.6 \LATITUDE -0.3
\RAHMS 13 58 00 \DECDM -62 09
\SIZE 8 \TYPE S
\FLUX1GHZ 5? \ALPHA ?
\NAMES Kes 20B

\RADIO Asymmetric shell.
\DATE 21 Aug 1996

\LONGITUDE 310.8 \LATITUDE -0.4
\RAHMS 14 00 00 \DECDM -62 17
\SIZE 12 \TYPE S
\FLUX1GHZ 6? \ALPHA ?
\NAMES Kes 20A

\RADIO Arc in E, in complex region.
\DATE 21 Aug 1996

\LONGITUDE 311.5 \LATITUDE -0.3
\RAHMS 14 05 38 \DECDM -61 58
\SIZE 5 \TYPE S
\FLUX1GHZ 3? \ALPHA 0.5
\NAMES

\RADIO Shell, not well resolved.
\DATE 21 Aug 1996

\LONGITUDE 312.4 \LATITUDE -0.4
\RAHMS 14 13 00 \DECDM -61 44
\SIZE 38 \TYPE S
\FLUX1GHZ 45 \ALPHA 0.36
\NAMES

\RADIO Irregular, incomplete shell.
\POINT Nearby $\gamma$-ray sources and pulsars.
\XRAY Weak emission in W.
\DISTANCE H\I/ absorption suggests $>6$~kpc and possibly $>14$~kpc.
\DATE 11 Jan 2004

\LONGITUDE 312.5 \LATITUDE -3.0
\RAHMS 14 21 00 \DECDM -64 12
\SIZE 18\X/20 \TYPE S
\FLUX1GHZ 3.5? \ALPHA ?
\NAMES

\RADIO Distorted shell.
\DATE 11 Jan 2004

\LONGITUDE 315.4 \LATITUDE -2.3
\RAHMS 14 43 00 \DECDM -62 30
\SIZE 42 \TYPE S
\FLUX1GHZ 49 \ALPHA 0.6
\NAMES RCW 86, MSH 14$-$6{\sl 3}

\NOTES Possibly the remnant of the SN of \AD/185?
\RADIO Shell, brightest to the SW.
\OPTICAL Bright, radiative filaments, with some faint Balmer dominated filaments.
\XRAY Partial shell, with thermal and non-thermal emission.
\POINT Several X-ray sources.
\DISTANCE Possible association with OB stars suggests 2.5~kpc.
\DATE 23 Jan 2004

\LONGITUDE 315.4 \LATITUDE -0.3
\RAHMS 14 35 55 \DECDM -60 36
\SIZE 24\X/13 \TYPE ?
\FLUX1GHZ 8 \ALPHA 0.4
\NAMES

\RADIO Irregular non-thermal emission, with H\II/ region superposed in E.
\DATE 20 Aug 1996

\LONGITUDE 315.9 \LATITUDE -0.0
\RAHMS 14 38 25 \DECDM -60 11
\SIZE 25\X/14 \TYPE S
\FLUX1GHZ 0.8? \ALPHA ?
\NAMES

\NOTES Has been called G315.8$-$0.0.
\RADIO Faint, distorted shell, with steep-spectrum `jet'?
\DATE 14 Aug 1998

\LONGITUDE 316.3 \LATITUDE -0.0
\RAHMS 14 41 30 \DECDM -60 00
\SIZE 29\X/14 \TYPE S
\FLUX1GHZ 20? \ALPHA 0.4
\NAMES (MSH 14$-$5{\sl 7})

\RADIO Distorted shell, with possible `blowout'.
\XRAY Detected.
\DISTANCE H\I/ absorption data suggests $>7.2$~kpc.
\DATE 10 Oct 2001

\LONGITUDE 317.3 \LATITUDE -0.2
\RAHMS 14 49 40 \DECDM -59 46
\SIZE 11 \TYPE S
\FLUX1GHZ 4.7? \ALPHA ?
\NAMES

\RADIO Incomplete shell.
\DATE 21 Aug 1996

\LONGITUDE 318.2 \LATITUDE +0.1
\RAHMS 14 54 50 \DECDM -59 04
\SIZE 40\X/35 \TYPE S
\FLUX1GHZ $>$3.9? \ALPHA ?
\NAMES

\RADIO Faint shell, with central H\II/ region.
\XRAY Sources within remnant.
\DATE 10 Oct 2001

\LONGITUDE 318.9 \LATITUDE +0.4
\RAHMS 14 58 30 \DECDM -58 29
\SIZE 30\X/14 \TYPE C
\FLUX1GHZ 4? \ALPHA 0.2?
\NAMES

\NOTES May not be a SNR?
\RADIO Complex arcs, with off-centre core.
\DATE 13 Aug 1998

\LONGITUDE 320.4 \LATITUDE -1.2
\RAHMS 15 14 30 \DECDM -59 08
\SIZE 35 \TYPE C
\FLUX1GHZ 60? \ALPHA 0.4
\NAMES MSH 15$-$5{\sl 2}, RCW 89

\NOTES Has been suggested as the remnant of the SN of \AD/185?
\RADIO Ragged shell.
\OPTICAL RCW 89 is the \Halpha/ emitting region to the NW.
\XRAY Partial shell, central nebula and pulsar and possible `jet'.
\POINT Radio and X-ray pulsar, with wind nebula.
\DISTANCE H\I/ absorption indicates 5.2~kpc.
\DATE 11 Jan 2004

\LONGITUDE 320.6 \LATITUDE -1.6
\RAHMS 15 17 50 \DECDM -59 16
\SIZE 60\X/30 \TYPE S
\FLUX1GHZ ? \ALPHA ?
\NAMES

\RADIO Faint shell, overlapping G320.4$-$1.2 in W.
\DATE 19 Aug 1998

\LONGITUDE 321.9 \LATITUDE -1.1
\RAHMS 15 23 45 \DECDM -58 13
\SIZE 28 \TYPE S
\FLUX1GHZ $>$3.4? \ALPHA ?
\NAMES

\RADIO Faint shell.
\DATE 13 Aug 1998

\LONGITUDE 321.9 \LATITUDE -0.3
\RAHMS 15 20 40 \DECDM -57 34
\SIZE 31\X/23 \TYPE S
\FLUX1GHZ 13 \ALPHA 0.3
\NAMES

\RADIO Shell brighter to the W, with Cir X-1 to N.
\POINT Compact, probably thermal source at S edge.
\DATE 7 Jan 2004

\LONGITUDE 322.5 \LATITUDE -0.1
\RAHMS 15 23 23 \DECDM -57 06
\SIZE 15 \TYPE C
\FLUX1GHZ 1.5 \ALPHA 0.4
\NAMES

\RADIO Shell with central extended source.
\POINT PN Pe 2-8 within boundary.
\DATE 13 Aug 1998

\LONGITUDE 323.5 \LATITUDE +0.1
\RAHMS 15 28 42 \DECDM -56 21
\SIZE 13 \TYPE S
\FLUX1GHZ 3? \ALPHA 0.4?
\NAMES

\RADIO Distorted shell, confused with thermal emission.
\POINT Compact, probably thermal source near centre.
\DATE 13 Aug 1998

\LONGITUDE 326.3 \LATITUDE -1.8
\RAHMS 15 53 00 \DECDM -56 10
\SIZE 38 \TYPE C
\FLUX1GHZ 145 \ALPHA varies
\NAMES MSH 15$-$5{\sl 6}

\NOTES Has been called G326.2$-$1.7.
\RADIO Shell, with elongated, flat-spectrum core.
\OPTICAL Emission around the shell.
\XRAY Shell, with central extended emission.
\DATE 10 Oct 2001

\LONGITUDE 327.1 \LATITUDE -1.1
\RAHMS 15 54 25 \DECDM -55 09
\SIZE 18 \TYPE C
\FLUX1GHZ 7? \ALPHA ?
\NAMES

\RADIO Shell, with off-centre core.
\XRAY Diffuse, with core.
\DATE 11 Jan 2004

\LONGITUDE 327.4 \LATITUDE +0.4
\RAHMS 15 48 20 \DECDM -53 49
\SIZE 21 \TYPE S
\FLUX1GHZ 30? \ALPHA 0.6
\NAMES Kes 27

\NOTES Has been called G327.3$+$0.4 and G327.3$+$0.5.
\RADIO Incomplete, multi-arc shell, brightest to the SE.
\XRAY Diffuse, best defined to E.
\DISTANCE H\I/ absorption indicates 4.3 to 5.4~kpc.
\DATE 11 Jan 2004

\LONGITUDE 327.4 \LATITUDE +1.0
\RAHMS 15 46 48 \DECDM -53 20
\SIZE 14 \TYPE S
\FLUX1GHZ 1.9? \ALPHA ?
\NAMES

\RADIO Asymmetric shell.
\DATE 10 Oct 2001

\LONGITUDE 327.6 \LATITUDE +14.6
\RAHMS 15 02 50 \DECDM -41 56
\SIZE 30 \TYPE S
\FLUX1GHZ 19 \ALPHA 0.6
\NAMES SN1006, PKS 1459$-$41

\NOTES This is the remnant of the SN of \AD/1006.
\RADIO Shell, with two bright arcs.
\OPTICAL Filaments to the NW, with broad \Halpha/ component.
\XRAY Thermal shell, with non-thermal limb-brightened arcs.
\POINT The background Schweizer--Middleditch star is near the middle of the remnant.
\DISTANCE Optical spectra and proper motion indicate $2.2\PM/0.1$~kpc.
\DATE 11 Jan 2004

\LONGITUDE 328.4 \LATITUDE +0.2
\RAHMS 15 55 30 \DECDM -53 17
\SIZE 5 \TYPE F
\FLUX1GHZ 15 \ALPHA 0.12
\NAMES (MSH 15$-$5{\sl 7})

\RADIO Amorphous emission, with central bar.
\XRAY Detected at high energies.
\DISTANCE H\I/ absorption indicates at least $17.4\PM/0.9$~kpc.
\DATE 10 Oct 2001

\LONGITUDE 329.7 \LATITUDE +0.4
\RAHMS 16 01 20 \DECDM -52 18
\SIZE 40\X/33 \TYPE S
\FLUX1GHZ $>$34? \ALPHA ?
\NAMES

\RADIO Diffuse shell, in complex region.
\DATE 10 Oct 2001

\LONGITUDE 330.0 \LATITUDE +15.0
\RAHMS 15 10 00 \DECDM -40 00
\SIZE 180? \TYPE S
\FLUX1GHZ 350? \ALPHA 0.5?
\NAMES Lupus Loop

\RADIO Low surface brightness loop with H\I/ shell.
\XRAY Detected.
\DATE 28 Jun 1995

\LONGITUDE 330.2 \LATITUDE +1.0
\RAHMS 16 01 06 \DECDM -51 34
\SIZE 11 \TYPE S?
\FLUX1GHZ 5? \ALPHA 0.3
\NAMES

\RADIO Clumpy non-thermal emission, possibly a distorted shell.
\DISTANCE H\I/ absorption indicates at least 4.9~kpc.
\DATE 10 Oct 2001

\LONGITUDE 332.0 \LATITUDE +0.2
\RAHMS 16 13 17 \DECDM -50 53
\SIZE 12 \TYPE S
\FLUX1GHZ 8? \ALPHA 0.5
\NAMES

\RADIO Incomplete shell.
\DATE 17 Oct 2001

\LONGITUDE 332.4 \LATITUDE -0.4
\RAHMS 16 17 33 \DECDM -51 02
\SIZE 10 \TYPE S
\FLUX1GHZ 28 \ALPHA 0.5
\NAMES RCW 103

\RADIO Shell, brightest to the S.
\OPTICAL Filaments correspond well to the radio shell, brightest in SE.
\XRAY Detected, with point source near centre.
\POINT Central, variable X-ray source, and nearby pulsar.
\DISTANCE H\I/ absorption indicates 3.3~kpc.
\DATE 7 Jan 2004

\LONGITUDE 332.4 \LATITUDE +0.1
\RAHMS 16 15 20 \DECDM -50 42
\SIZE 15 \TYPE S
\FLUX1GHZ 26 \ALPHA 0.5
\NAMES MSH 16$-$5{\sl 1}, Kes 32

\NOTES Has been called G332.4$+$0.2.
\RADIO Distorted shell, with thermal jet and plume adjacent.
\POINT Pulsar nearby.
\DATE 7 Jan 2004

\LONGITUDE 335.2 \LATITUDE +0.1
\RAHMS 16 27 45 \DECDM -48 47
\SIZE 21 \TYPE S
\FLUX1GHZ 16 \ALPHA 0.5
\NAMES

\RADIO Well defined shell.
\POINT Old pulsar within remnant boundary.
\DATE 13 Aug 1998

\LONGITUDE 336.7 \LATITUDE +0.5
\RAHMS 16 32 11 \DECDM -47 19
\SIZE 14\X/10 \TYPE S
\FLUX1GHZ 6 \ALPHA 0.5
\NAMES

\RADIO Irregular shell.
\DATE 13 Aug 1998

\LONGITUDE 337.0 \LATITUDE -0.1
\RAHMS 16 35 57 \DECDM -47 36
\SIZE 1.5 \TYPE S
\FLUX1GHZ 1.5 \ALPHA 0.6?
\NAMES (CTB 33)

\NOTES This entry refers to a small ($1'.5$) SNR, not the larger previously catalogued G337.0$-$0.1.
\RADIO Shell, in a complex region.
\DISTANCE Association with CTB 33 gives 11~kpc.
\POINT Associated with a soft gamma repeater.
  lines, clarifying extent of the remnant.
\DATE 17 Oct 2001

\LONGITUDE 337.2 \LATITUDE -0.7
\RAHMS 16 39 28 \DECDM -47 51
\SIZE 6 \TYPE S
\FLUX1GHZ 2? \ALPHA 0.7
\NAMES

\RADIO Shell, not well resolved.
\XRAY Extended emission.
\DATE 10 Oct 2001

\LONGITUDE 337.3 \LATITUDE +1.0
\RAHMS 16 32 39 \DECDM -46 36
\SIZE 15\X/12 \TYPE S
\FLUX1GHZ 16 \ALPHA 0.55
\NAMES Kes 40

\RADIO Nearly complete shell.
\DATE 13 Aug 1998

\LONGITUDE 337.8 \LATITUDE -0.1
\RAHMS 16 39 01 \DECDM -46 59
\SIZE 9\X/6 \TYPE S
\FLUX1GHZ 18 \ALPHA 0.5
\NAMES Kes 41

\RADIO Distorted shell.
\DISTANCE H\I/ absorption suggests $>9.3$~kpc.
\DATE 1 Aug 2000

\LONGITUDE 338.1 \LATITUDE +0.4
\RAHMS 16 37 59 \DECDM -46 24
\SIZE 15? \TYPE S
\FLUX1GHZ 4? \ALPHA 0.4
\NAMES

\RADIO Arc in NE, merging with thermal emission in S.
\OPTICAL Detected.
\XRAY Detected.
\DATE 10 Oct 2001

\LONGITUDE 338.3 \LATITUDE -0.0
\RAHMS 16 41 00 \DECDM -46 34
\SIZE 8 \TYPE S
\FLUX1GHZ 7? \ALPHA ?
\NAMES

\RADIO Irregular shell, in complex region.
\DATE 13 Aug 1998

\LONGITUDE 338.5 \LATITUDE +0.1
\RAHMS 16 41 09 \DECDM -46 19
\SIZE 9 \TYPE ?
\FLUX1GHZ 12? \ALPHA ?
\NAMES

\RADIO Circle of non-thermal emission in complex region, not well defined.
\DATE 13 Aug 1998

\LONGITUDE 340.4 \LATITUDE +0.4
\RAHMS 16 46 31 \DECDM -44 39
\SIZE 10\X/7 \TYPE S
\FLUX1GHZ 5 \ALPHA 0.4
\NAMES

\RADIO Distorted shell, elongated east--west.
\DATE 13 Aug 1998

\LONGITUDE 340.6 \LATITUDE +0.3
\RAHMS 16 47 41 \DECDM -44 34
\SIZE 6 \TYPE S
\FLUX1GHZ 5? \ALPHA 0.4?
\NAMES

\RADIO Incomplete shell.
\OPTICAL Possible associated filaments.
\DATE 1 Aug 2000

\LONGITUDE 341.2 \LATITUDE +0.9
\RAHMS 16 47 35 \DECDM -43 47
\SIZE 16\X/22 \TYPE C
\FLUX1GHZ 1.5? \ALPHA 0.6?
\NAMES

\RADIO Incomplete shell, with extension to SW.
\POINT Pulsar in W, with wind nebula.
\DATE 8 Nov 2001

\LONGITUDE 341.9 \LATITUDE -0.3
\RAHMS 16 55 01 \DECDM -44 01
\SIZE 7 \TYPE S
\FLUX1GHZ 2.5 \ALPHA 0.5
\NAMES

\RADIO Incomplete shell, brightest to NE.
\DATE 27 Jan 2004

\LONGITUDE 342.0 \LATITUDE -0.2
\RAHMS 16 54 50 \DECDM -43 53
\SIZE 12\X/9 \TYPE S
\FLUX1GHZ 3.5? \ALPHA 0.4?
\NAMES

\RADIO Distorted shell.
\DATE 1 Aug 2000

\LONGITUDE 342.1 \LATITUDE +0.9
\RAHMS 16 50 43 \DECDM -43 04
\SIZE 10\X/9 \TYPE S
\FLUX1GHZ 0.5? \ALPHA ?
\NAMES

\RADIO Incomplete shell.
\DATE 13 Aug 1998

\LONGITUDE 343.0 \LATITUDE -6.0
\RAHMS 17 25 00 \DECDM -46 30
\SIZE 250 \TYPE S
\FLUX1GHZ ? \ALPHA ?
\NAMES

\RADIO Faint, poorly defined.
\OPTICAL Filamentary shell.
\DATE 7 Jan 2004

\LONGITUDE 343.1 \LATITUDE -2.3
\RAHMS 17 08 00 \DECDM -44 16
\SIZE 32? \TYPE C?
\FLUX1GHZ 8? \ALPHA 0.5?
\NAMES

\RADIO Incomplete shell?
\XRAY Pulsar wind nebula.
\POINT Pulsar near edge, with wind nebula.
\DATE 27 Jan 2004

\LONGITUDE 343.1 \LATITUDE -0.7
\RAHMS 17 00 25 \DECDM -43 14
\SIZE 27\X/21 \TYPE S
\FLUX1GHZ 7.8 \ALPHA 0.55
\NAMES

\RADIO Shell, with smaller thermal shell adjacent.
\DATE 1 Aug 2000

\LONGITUDE 344.7 \LATITUDE -0.1
\RAHMS 17 03 51 \DECDM -41 42
\SIZE 10 \TYPE C?
\FLUX1GHZ 2.5? \ALPHA 0.5
\NAMES

\RADIO Aysmmetric shell, with possible core.
\XRAY Detected.
\DATE 10 Oct 2001

\LONGITUDE 345.7 \LATITUDE -0.2
\RAHMS 17 07 20 \DECDM -40 53
\SIZE 6 \TYPE S
\FLUX1GHZ 0.6? \ALPHA ?
\NAMES

\RADIO Poorly defined diffuse shell.
\POINT Old pulsar nearby.
\DATE 13 Aug 1998

\LONGITUDE 346.6 \LATITUDE -0.2
\RAHMS 17 10 19 \DECDM -40 11
\SIZE 8 \TYPE S
\FLUX1GHZ 8? \ALPHA 0.5?
\NAMES

\RADIO Irregular shell.
\DATE 1 Aug 2000

\LONGITUDE 347.3 \LATITUDE -0.5
\RAHMS 17 13 50 \DECDM -39 45
\SIZE 65\X/55 \TYPE S?
\FLUX1GHZ ? \ALPHA ?
\NAMES

\RADIO Faint emission.
\XRAY Non-thermal, limb-brightended to W, with central source.
\POINT Central X-ray source.
\DISTANCE Association with molecular clouds, and H\II/ region, suggests 1~kpc or 6~kpc.
\DATE 23 Jan 2004

\LONGITUDE 348.5 \LATITUDE -0.0
\RAHMS 17 15 26 \DECDM -38 28
\SIZE 10? \TYPE S?
\FLUX1GHZ 10? \ALPHA 0.4?
\NAMES

\RADIO Arc, overlapping G348.5$+$0.1.
  ($2''.5\X/3''.9$).
\DATE 1 Aug 2000

\LONGITUDE 348.5 \LATITUDE +0.1
\RAHMS 17 14 06 \DECDM -38 32
\SIZE 15 \TYPE S
\FLUX1GHZ 72 \ALPHA 0.3
\NAMES CTB 37A

\RADIO Shell, poorly define to S and W, overlapping G348.5$-$0.0 in E.
\DISTANCE H\I/ absorption indicates 10.2$\PM/$3.5~kpc.
\DATE 7 Jan 2004

\LONGITUDE 348.7 \LATITUDE +0.3
\RAHMS 17 13 55 \DECDM -38 11
\SIZE 17? \TYPE S
\FLUX1GHZ 26 \ALPHA 0.3
\NAMES CTB 37B

\RADIO Incomplete shell with faint eastern extensions.
\DISTANCE H\I/ absorption indicates 10.2$\PM/$3.5~kpc.
\DATE 21 Aug 1996

\LONGITUDE 349.2 \LATITUDE -0.1
\RAHMS 17 17 15 \DECDM -38 04
\SIZE 9\X/6 \TYPE S
\FLUX1GHZ 1.4? \ALPHA ?
\NAMES

\RADIO Elongated shell, adjacent to bright H\II/ region.
\DATE 27 Aug 1996

\LONGITUDE 349.7 \LATITUDE +0.2
\RAHMS 17 17 59 \DECDM -37 26
\SIZE 2.5\X/2 \TYPE S
\FLUX1GHZ 20 \ALPHA 0.5
\NAMES

\RADIO Incomplete clumpy shell, with enhancement to the S.
\DISTANCE H\I/ absorption indicates $18.3\PM/4.6$~kpc, association with OH features gives 22~kpc.
\XRAY Irregular shell.
\DATE 23 Jan 2004

\LONGITUDE 350.0 \LATITUDE -2.0
\RAHMS 17 27 50 \DECDM -38 32
\SIZE 45 \TYPE S
\FLUX1GHZ 26 \ALPHA 0.4
\NAMES

\NOTES Incorporates the previously catalogued G350.0$-$1.8 in the NW.
\RADIO Shell, brightest in NW.
\DATE 13 Aug 1998

\LONGITUDE 351.2 \LATITUDE +0.1
\RAHMS 17 22 27 \DECDM -36 11
\SIZE 7 \TYPE C?
\FLUX1GHZ 5? \ALPHA 0.4
\NAMES

\NOTES Has been called G351.3$+$0.2.
\RADIO Distorted shell, with possible flat-spectrum core.
\DATE 13 Aug 1998

\LONGITUDE 351.7 \LATITUDE +0.8
\RAHMS 17 21 00 \DECDM -35 27
\SIZE 18\X/14 \TYPE S
\FLUX1GHZ 10? \ALPHA ?
\NAMES

\RADIO Elongated shell, adjacent to bright H\II/ region.
\POINT Pulsar nearby.
\DATE 21 Aug 1996

\LONGITUDE 351.9 \LATITUDE -0.9
\RAHMS 17 28 52 \DECDM -36 16
\SIZE 12\X/9 \TYPE S
\FLUX1GHZ 1.8? \ALPHA ?
\NAMES

\RADIO Asymmetric shell.
\DATE 21 Aug 1996

\LONGITUDE 352.7 \LATITUDE -0.1
\RAHMS 17 27 40 \DECDM -35 07
\SIZE 8\X/6 \TYPE S
\FLUX1GHZ 4 \ALPHA 0.6
\NAMES

\RADIO Distorted shell.
\XRAY Detected.
\DATE 10 Oct 2001

\LONGITUDE 353.9 \LATITUDE -2.0
\RAHMS 17 38 55 \DECDM -35 11
\SIZE 13 \TYPE S
\FLUX1GHZ 1? \ALPHA 0.5?
\NAMES

\RADIO Shell, with central double source.
\DATE 8 Nov 2001

\LONGITUDE 354.1 \LATITUDE +0.1
\RAHMS 17 30 28 \DECDM -33 46
\SIZE 15\X/3? \TYPE C?
\FLUX1GHZ ? \ALPHA varies?
\NAMES

\NOTES Is this a SNR?
\RADIO Elongated N--S.
\POINT Pulsar at S tip.
\DATE 31 Jul 1995

\LONGITUDE 354.8 \LATITUDE -0.8
\RAHMS 17 36 00 \DECDM -33 42
\SIZE 19 \TYPE S
\FLUX1GHZ 2.8? \ALPHA ?
\NAMES

\RADIO Distorted shell.
\DATE 1 Aug 2000

\LONGITUDE 355.6 \LATITUDE -0.0
\RAHMS 17 35 16 \DECDM -32 38
\SIZE 8\X/6 \TYPE S
\FLUX1GHZ 3? \ALPHA ?
\NAMES

\RADIO Well defined shell.
\XRAY Detected.
\DATE 10 Oct 2001

\LONGITUDE 355.9 \LATITUDE -2.5
\RAHMS 17 45 53 \DECDM -33 43
\SIZE 13 \TYPE S
\FLUX1GHZ 8 \ALPHA 0.5
\NAMES

\RADIO Distorted shell, brightest to SE.
\DATE 13 Aug 1998

\LONGITUDE 356.2 \LATITUDE +4.5
\RAHMS 17 19 00 \DECDM -29 40
\SIZE 25 \TYPE S
\FLUX1GHZ 4 \ALPHA 0.7
\NAMES

\RADIO Faint shell.
\DATE 12 Oct 2001

\LONGITUDE 356.3 \LATITUDE -0.3
\RAHMS 17 37 56 \DECDM -32 16
\SIZE 11\X/7 \TYPE S
\FLUX1GHZ 3? \ALPHA ?
\NAMES

\RADIO Elongated shell, brighter in N.
\DATE 11 Jan 2004

\LONGITUDE 356.3 \LATITUDE -1.5
\RAHMS 17 42 35 \DECDM -32 52
\SIZE 20\X/15 \TYPE S
\FLUX1GHZ 3? \ALPHA ?
\NAMES

\RADIO Double arc.
\DATE 30 Jun 1995

\LONGITUDE 357.7 \LATITUDE -0.1
\RAHMS 17 40 29 \DECDM -30 58
\SIZE 8\X/3? \TYPE ?
\FLUX1GHZ 37 \ALPHA 0.4
\NAMES MSH 17$-$3{\sl 9}

\NOTES Has been suggested that this is not a SNR.
\RADIO Multiple arcs and filaments, with compact H\II/ region at W edge.
\XRAY Detected.
\DISTANCE H\I/ absorption suggests beyond Galactic Centre.
\DATE 23 Jan 2004

\LONGITUDE 357.7 \LATITUDE +0.3
\RAHMS 17 38 35 \DECDM -30 44
\SIZE 24 \TYPE S
\FLUX1GHZ 10 \ALPHA 0.4?
\NAMES

\RADIO Non-thermal shell in complex region.
  from surveys.
\DATE 1 Aug 2000

\LONGITUDE 358.0 \LATITUDE +3.8
\RAHMS 17 26 00 \DECDM -28 36
\SIZE 38 \TYPE S
\FLUX1GHZ 1.5? \ALPHA ?
\NAMES

\RADIO Faint shell.
\DATE 12 Oct 2001

\LONGITUDE 359.0 \LATITUDE -0.9
\RAHMS 17 46 50 \DECDM -30 16
\SIZE 23 \TYPE S
\FLUX1GHZ 23 \ALPHA 0.5
\NAMES

\RADIO Incomplete shell.
\XRAY Partial shell.
\DATE 1 Aug 2000

\LONGITUDE 359.1 \LATITUDE -0.5
\RAHMS 17 45 30 \DECDM -29 57
\SIZE 24 \TYPE S
\FLUX1GHZ 14 \ALPHA 0.4?
\NAMES

\RADIO Non-thermal shell in complex region, crossed by the `snake'.
\XRAY Centrally brightened.
\POINT Several compact radio sources near centre, OH masers around edge.
\DATE 7 Jan 2004

\LONGITUDE 359.1 \LATITUDE +0.9
\RAHMS 17 39 36 \DECDM -29 11
\SIZE 12\X/11 \TYPE S
\FLUX1GHZ 5? \ALPHA ?
\NAMES

\RADIO Shell, brightest in E.
\DATE 1 Aug 2000


\centerline{\hrulefill}
\label{lastpage}

\end{document}